\documentclass[11pt,a4paper]{article}
\usepackage{jstyle}
\usepackage{hyperref,slashed,empheq}
\usepackage{framed}
\usepackage{caption}
\usepackage{subcaption}
\usetikzlibrary{calc,decorations.markings}
\newcommand{\Y}{\mathbb{Y}}
\newcommand{\N}{\mathbb{N}}

\newcommand{\dd}{\mathrm{d}}

\newcommand{\Pd}[1]{\mathcal{P}_{#1}}
\newcommand{\rac}{\mathrm{Rac}}
\newcommand{\di}{\mathrm{Di}}
\newcommand{\half}{\boldsymbol{\tfrac12}}

\definecolor{mygreen}{rgb}{0.32,1.65,0.50}
\definecolor{myblue}{rgb}{0.08, 0.38, 0.74}
\definecolor{theRed}{rgb}{0.56,0,0}

\author[a]{Thomas BASILE}
\author[a]{\quad Euihun JOUNG}
\author[b]{\quad Shailesh LAL}
\author[a]{\quad Wenliang LI}

\affiliation[a]{Department of Physics and Research Institute of Basic
  Science, \\ Kyung Hee University,\\ Seoul 02447, Korea}
\affiliation[b]{Centro de Fisica do Porto e Departamento de Fisica e
  Astronomia da\\ Faculdade de Ciencias da Universidade do
  Porto,\\ Rua do Campo Alegre 687, 4169-007 Porto, Portugal}

\emailAdd{thomas.basile@khu.ac.kr}
\emailAdd{euihun.joung@khu.ac.kr}
\emailAdd{slal@fc.up.pt}
\emailAdd{lii.wenliang@gmail.com}

\abstract{We compute the one-loop free energies of the type-A$_\ell$
  and type-B$_\ell$ higher-spin gravities in $(d+1)$-dimensional
  anti-de Sitter (AdS$_{d+1}$) spacetime. For large $d$ and $\ell$,
  these theories have a complicated field content, and hence it is
  difficult to compute their zeta functions using the usual methods.
  Applying the character integral representation of zeta function
  developed in the companion paper
  \href{https://arxiv.org/abs/1805.05646}{\tt [1805.05646]} to these
  theories, we show how the computation of their zeta function can be
  shortened considerably.  We find that the results previously
  obtained for the massless theories ($\ell=1$) generalize to their
  partially-massless counterparts (arbitrary $\ell$) in arbitrary
  dimensions.
}

\begin{document}

\title{\centering {\huge Character Integral Representation\\ of Zeta
    function in AdS$_{d+1}$}:\\
  \bigskip
  \Large II. Application to partially-massless higher-spin gravities}

\maketitle

\section{Introduction}
\label{sec:intro}
Holographic dualities involving higher-spin  gravities in
AdS$_{d+1}$ and vector model Conformal Field Theories (CFTs) on its
$d$-dimensional boundary have been explored on a variety of fronts. As
is well known by now, the single trace sector in the large-$N$
expansion of free vector models in $d$ dimensions with Lagrangian
densities
\begin{equation}\label{masslesscft}
  \mathcal{L} = \phi^{*i}\, \Box\,\phi_i\quad \text{and}\quad
  \mathcal{L} = \bar{\psi}^i\, \slashed{\partial}\,\psi_i\,,
\end{equation}
are respectively dual to the type-A
\cite{Vasiliev:1990en,Vasiliev:1992av,Vasiliev:2003ev} and type-B
Vasiliev theories in AdS$_{d+1}$ \cite{Klebanov:2002ja,
  Sezgin:2002rt}.  Here $\phi$ is a complex scalar, $\psi$ is a Dirac
spinor, and the index $i$ is a vector index of $U(N)$ (i.e. both
fields are in the fundamental representation of $U(N)$). If we
restrict to real scalars and Majorana fermions, $U(N)$ is replaced by
$O(N)$ and the AdS dual is the minimal type-A and type-B theory.  In
AdS$_4$, one can also consider the large $N$ limit of the critical
$O(N)$ model, which is obtained by a double trace deformation
\cite{Klebanov:2002ja,Gubser:2002zh}. We refer the reader to
\cite{Giombi:2012ms, Giombi:2016ejx, Sleight:2016hyl, Sleight:2017krf}
for reviews of the duality.

It turns out that if we relax the criterion of unitarity, it is
natural to consider the following one-parameter extension of the CFTs
in \eqref{masslesscft}, given by
\begin{equation}\label{pmcftdual}
  \mathcal{L} = \phi^{*i}\, \Box\,^\ell\,\phi_i\quad \text{and}\quad
  \mathcal{L} = \bar{\psi}^i\,
  \slashed{\partial}\,^{2\ell-1}\,\psi_i\,.
\end{equation}
It was conjectured in \cite{Bekaert:2013zya} for the bosonic case that
\eqref{pmcftdual} is the CFT dual of an interacting AdS theory
containing both massless and partially-massless higher-spin fields
\cite{Deser:1983mm,Higuchi:1986wu,Deser:2001us} (which should be dual
to partially-conserved currents \cite{Dolan:2001ih}). On the one hand,
the bulk side of this duality corresponds to the partially-massless
higher-spin gravity, which is also referred to as the type-A$_\ell$
theory. Cubic interactions for partially-massless field were derived
in the metric-like formulation in \cite{Joung:2012hz,
  Joung:2012rv}\,\footnote{See also \cite{Joung:2014aba} concerning
  the non-unitary nature of an interacting theory for the partially
  massless spin-$2$ field.}  whereas the unfolded equations for the
type-A$_\ell$ theory were constructed first in
\cite{Alkalaev:2014nsa}, and recently studied in more details in
\cite{Brust:2016zns} for $\ell=2$. On the other hand, the CFT on the
boundary was discussed in \cite{Brust:2016gjy, Gubser:2017vgc} (see
also \cite{Safari:2017tgs, Safari:2017irw} for a more detailed study
of its critical counterparts). The symmetry algebra underlying the
kinematics of this correspondence was analyzed in
\cite{Bekaert:2013zya, Alkalaev:2014nsa, Joung:2015jza}, whereas its
Eastwood-like characterization was provided in \cite{Eastwood2008,
  Michel2011, Gover2012}. For the fermionic vector models
\eqref{pmcftdual}, the putative dual theories are the type-B$_\ell$
gravities about which much less is known.  For instance, a set of
formal non-linear equations was proposed only recently for the
massless $(\ell=1)$ case in \cite{Grigoriev:2018wrx}.

Let us briefly review the systematics of testing the duality for the
one-loop free energy,\,\footnote{See  
    \cite{Sezgin:2003pt, Giombi:2009wh, Giombi:2010vg, Colombo:2012jx,
      Didenko:2012tv,Bonezzi:2017vha} for other tests of the
    duality between Vasiliev's type-A theory and the free $O(N)$
    vector model.
    See  \cite{Bekaert:2014cea,Bekaert:2015tva,Bekaert:2016ezc,Sleight:2016dba,Sleight:2017fpc,Ponomarev:2017qab} 
    for the attempts of reconstructing higher-spin gravity action from the CFT data.}
    following the
arguments of \cite{Giombi:2013fka,Giombi:2014iua}. For definiteness we
focus on the type-A$_\ell$ theories, but the same arguments apply to
the type-B$_\ell$ case. The free energy of the CFT$_d$ is simply given
by
\begin{equation}\label{fcft}
  F_{\rm\sst CFT} = k\,N\,F_{\ell},
\end{equation}
where $F_{\ell}$ is the free energy of the free $2\,\ell$-derivatives
scalar theory,\,\footnote{The scalar field of this class of free
  theories will be referred to, in the rest of the paper, as the
  order-$\ell$ scalar singleton.} and $k=2$ for the $U(N)$ vector
model and $k=1$ for the $O(N)$ vector model.  For even $d$, by the
free energy of CFT, we actually mean the $a$-anomaly coefficient.
Meanwhile on the AdS$_{d+1}$ side the free energy has the expansion
around the AdS saddle point
\begin{equation}
  F_{\rm\sst AdS} = g^{-1}\,\Gamma^{\sst (0)} + \Gamma^{\sst
    (1)}+\ldots\,,
\end{equation}
where $\Gamma^{\sst (0)}$ and $\Gamma^{\sst (1)}$ are respectively the
renormalized\,\footnote{The bulk free energy is divergent due to the
  infinite volume of AdS. However the divergence may be renormalized
  in accordance with general principles of AdS/CFT
  duality. Practically, this requires us to replace the infinite
  volume of AdS that appears in the free energy with a well-known
  finite quantity.} semi-classical and one-loop contributions to the
AdS free energy and $g$ is the bulk coupling constant. Since the
AdS/CFT dictionary indicates $g^{-1}\sim N$, requiring that
$F_{\rm\sst AdS}=F_{\rm\sst CFT}$ leads us to expect that {\it i)} the
background evaluation of the type-A$_\ell$ higher-spin gravity action
should completely reproduce $F_{\rm\sst CFT}$, and {\it ii)} the
one-loop free energy of the type-A$_\ell$ higher-spin gravity, which
corresponds to the (vanishing) $N^0$ contribution in $F_{\rm\sst CFT}$
of the dual free CFT, should simply vanish.  Since we do not know the
classical action of the higher-spin gravity, we cannot test the first
point, but the second point about the one-loop free energy can be
examined.  Besides the usual UV divergence, the one-loop free energy
of higher-spin gravity has another source of divergence arising from
summing over an infinite number of particles in the spectrum. This may
be regularized in various ways \cite{Giombi:2013fka, Giombi:2014iua,
  Giombi:2014yra} (see also \cite{Tseytlin:2013jya} in the context of
conformal higher-spin), among which the zeta function regularization
was particularly appealing as the UV regulator turns out to regularize
the divergence from the infinite spectrum as well.  In the $\ell=1$
case, it was found in \cite{Giombi:2013fka,Giombi:2014iua} that the
one-loop free energy of the non-minimal theory indeed vanishes.
However, the result of the minimal theory does not vanish, giving a
number which coincides with the free energy of the real scalar on the
$S^d$ boundary.  This result was interpreted as an indication that the
relation between the bulk coupling constant $g$ and the boundary $N$
should be modified to $g^{-1} = N-1$\,.  Then the sum of the
semi-classical and the one-loop contributions match the CFT free
energy.

It is tempting to expect that similar statements would hold for the
$\ell\ge 2$ cases. Indeed, the computations for $\ell=2$ carried out
for various values of $d$ (up to $d=18$ and $d=7$ for even and odd
$d$, respectively, in \cite{Brust:2016xif}) seem to support this
expectation.  However, testing it for a larger $\ell$ becomes highly
non-trivial because the field content itself becomes increasingly
complicated as $\ell$ grows.  For instance, the field content of the
type-A$_{\ell=3}$ higher-spin gravity involves three series:
\begin{equation}
  {\rm A}_{\ell=3}^{\text{non-min}}\cong \bigoplus_{s=0}^\infty \,
  \mathcal{D}\big(s+d-2; s\big) \oplus\bigoplus_{s=0}^\infty \,
  \mathcal{D}\big(s+d-4; s\big) \oplus\bigoplus_{s=0}^\infty \,
  \mathcal{D}\big(s+d-6; s\big)\,.
  \label{A ex}
\end{equation}
About the type-B$_\ell$ theory, already the $\ell=1$ case has rather
complicated spectrum as it starts to involve mixed-symmetry fields in
$d\ge 4$. In spite of the complexity of the field content, its
one-loop free energy has been computed up to 
$d=20$ and $d=15$ for even and odd $d$, respectively
\cite{Giombi:2013fka, Giombi:2014iua, Gunaydin:2016amv,
  Giombi:2016pvg}, confirming the aforementionned
expectations.\,\footnote{The test of higher-spin holographic dualities was also
  extended to the type-C theory \cite{Beccaria:2014xda,
    Beccaria:2014zma, Beccaria:2014qea,Giombi:2016pvg}, but we will
  not address this case in this paper.} However, if we consider higher
$\ell$'s and the minimal theory, then the spectrum of the
type-B$_\ell$ theory becomes almost untreatable: for general $\ell$
and $d$, the spectrum does not have a simple form but a rather lengthy
expression which can be found in \hyperref[app:minBl]{Appendix
  \ref{app:minBl}}. For instance, the minimal type-B$_{\ell=2}$ theory
spectrum reads
\begin{eqnarray}
  {\rm B}_{\ell=2}^{\rm min} \quad & \cong & \quad
  \mathcal{D}\big(d-3;0\big) \oplus \mathcal{D}\big(d-1;0\big) \oplus
  \mathcal{D}\big(d;0\big) \oplus \mathcal{D}\big(d+1;0\big) \\ &&
  \qquad \oplus \bigoplus_{m=1,2\,\,{\rm mod}\, 4}\,\,
  \bigoplus_{s=2,4,\dots}^\infty 2\, \mathcal{D}\big( s+d-2;s,1^m\big)
  \oplus \mathcal{D}\big( s+d-4;s,1^m\big) \nn && \qquad \oplus
  \bigoplus_{m=0,3\,\,{\rm mod}\, 4}\,\,
  \bigoplus_{s=1,3,\dots}^\infty 2\, \mathcal{D}\big( s+d-2;s,1^m\big)
  \oplus \mathcal{D}\big( s+d-4;s,1^m\big) \nn && \qquad \oplus
  \bigoplus_{m=0}^{r-1}\,\, \bigoplus_{s=1}^\infty \mathcal{D}\big(
  s+d-3;s,1^m\big)\,, \nonumber
  \label{B ex}
\end{eqnarray}
for $d=5$ mod $8$. To reiterate, as one can see from the field
contents \eqref{A ex} and \eqref{B ex}, computing the one-loop free
energy for arbitrary $\ell$ and $d$ using the usual methods of
spectrum summations would be almost impossible.

In this paper, we apply the method of the character integral
representation of zeta function (CIRZ) to calculate the one-loop free
energy of the partially-massless higher-spin gravities in AdS$_{d+1}$,
and match the result with the free energy of the corresponding CFT on
the $S^d$ boundary.  The CIRZ was originally devised in
\cite{Bae:2016rgm} to study the one-loop free energy of a stringy
theory in AdS$_4$ and AdS$_5$ dual to free matrix CFTs. It proved
useful in several related applications \cite{Bae:2016hfy, Bae:2017fcs,
  Pang:2016ofv, Bae:2017spv} and generalized to arbitrary dimensions
in the companion paper \cite{Basile:2018zoy}.  In the latter paper,
the CIRZ was obtained as a contour integrals of the character of the
representation underlying the AdS theory.  This contour integral
expression allows us to handle the dependence on $d$ and $\ell$ of the
partially-massless higher-spin gravity in an analytic manner.
Moreover, both the AdS and CFT quantities reduce to a compact integral
and the match can be demonstrated at the level of the integral,
thereby extending the result of \cite{Skvortsov:2017ldz} dedicated to
the type-A$_{\ell=1}$ theory to that of $\ell>1$. As a result, we
provide a test of the type-A$_\ell$ and type-B$_\ell$ dualities for
all $d$ and $\ell$.

The organization of the paper is as follows. In \hyperref[sec:
  CIRZ]{Section \ref{sec: CIRZ}}, we give a brief overview of the CIRZ
method derived in \cite{Basile:2018zoy}. In
\hyperref[sec:pmhsg]{Section \ref{sec:pmhsg}}, we then turn to a
review of some key facts about the field content of type-A$_\ell$ and
type-B$_\ell$ theories. We next apply the CIRZ method to type-A$_\ell$
and type-B$_\ell$ theories to compute the one-loop free energy in
\hyperref[sec:application]{Section \ref{sec:application}} and
\hyperref[sec:typeBl]{Section \ref{sec:typeBl}}, while a few
generalizations of these theories are considered in
\hyperref[sec:genhigher-spin]{Section \ref{sec:genhigher-spin}}. We
finally conclude with a discussion of our results.
\hyperref[app:minBl]{Appendix \ref{app:minBl}} contains the explicit
spectrum of the minimal type-B$_\ell$ theory.

\section{CIRZ Formula and One-Loop Free Energy}
\label{sec: CIRZ}
In this section we will briefly recollect the CIRZ formulas obtained
in the companion paper \cite{Basile:2018zoy}. We shall mainly focus on
the application of these formulas to extracting the one-loop free
energy of the AdS$_{d+1}$ theory.

\subsection{CIRZ in arbitrary dimensions}
To recapitulate briefly, the contribution of a given particle,
carrying an $so(2,d)$ irrep $[\D;\Y]$,\,\footnote{A field in
  AdS$_{d+1}$ is labelled by a lowest weight $[\Delta;\mathbb{Y}]$ of
  $so(2,d)$, the isometry algebra of spacetime. Here $\Delta$ denotes
  the minimum energy (or the conformal dimension of the dual CFT$_{d}$
  operator), and $\mathbb{Y} = (s_1,\ldots,s_r)$ with $r=[\tfrac d2]$
  (where $[x]$ denotes the integer part of $x$) is an $so(d)$ lowest
  weight, i.e. the `spin' of the field.} to the one-loop free energy
of the theory is given in terms of the spectral zeta function
$\zeta_{[\D;\Y]}(z)$ as
\begin{equation}\label{free energy}
  \Gamma^{\sst (1)}_{[\D;\Y]}=-\frac\e2\,
  \zeta_{[\Delta;\mathbb{Y}]}(0) \ln(R\,\Lambda_{\rm\sst
    UV})-\frac\e2\,\zeta'_{[\Delta;\mathbb{Y}]}(0)\,,
\end{equation}
where $\e$ is the sign $+/-$ for boson/fermion and $R$ and
$\Lambda_{\rm\sst UV}$ are the AdS radius and an ultraviolet (UV)
cutoff, respectively.  An integral representation of the zeta function
$\zeta_{[\Delta;\mathbb{Y}]}(z)$ is derived by Camporesi and Higuchi
in \cite{Camporesi1994}. The CIRZ reformulates this zeta function as
an integral transform of the character $\chi^{so(2,d)}_{(\Delta;
  \Y)}(\beta, \vec\alpha)$. In this way, the CIRZ allows to sum the
zeta functions over fields in a theory using the corresponding
characters. If an AdS theory has a field content carrying a
\emph{reducible} representation $\cH$ of the isometry algebra
$so(2,d)$, then the zeta function of the theory is given as follows:
when $d=2r$, it is
\begin{equation}
  \zeta_{\cH}(z) = \ln R\, \int_0^\infty \frac{\dd
    \beta}{\G(z)^2}\,\left(\frac\b 2\right)^{2(z-1)}\,
  f_{\cH}(z,\b),
  \label{GZf}
\end{equation}
with
\begin{equation}
  f_{\cH}(z,\b) =\sum_{k=0}^r\oint_C \mu(\bm \a)\,
  \left(1+\big(\tfrac{\a_k}{\b}\big)^2\right)^{z-1}\,
  \prod_{\substack{0 \leqslant j \leqslant r \\ j \neq k}}
  \frac{\cosh\beta-\cos\alpha_j}{\cos\alpha_k-\cos\alpha_j}\,
  \chi^{so(2,d)}_{\cH}(\beta; \vec \alpha_k)\,.
  \label{f even}
\end{equation}
We can use this expression to prove, for example, that
$\zeta_{[\Delta;\Y]}(0)$ vanishes identically, which corresponds to
the well-known absence of logarithmic divergences in AdS$_{2r+1}$ free
energy. This is due to the presence of the $1\over\Gamma(z)^2$ factor
in the expression for $\zeta_{\cH}(z)$ above. When $d=2r+1$, the
primary contribution of the zeta function is given by
\begin{equation}
  \zeta_{1,\cH}(z) = \int_0^\infty \frac{\dd
    \beta\,\b^{2z-1}}{\Gamma(2z)}\, f_{1,\cH}(\b),
  \label{zeta 1 even AdS}
\end{equation}
with
\begin{equation}
  \begin{split}
    f_{1,\cH}(\b) =\sum_{k=0}^r\,\oint_C \mu(\bm \a)&\,
    \frac{\sinh\frac\b2\,(\cosh\frac\b2)^\frac{1+\e}2
      (\cos\frac{\a_k}2)^\frac{1-\e}2}{\cosh\b-\cos\a_k} \\&\qquad
    \times\, \prod_{\substack{0 \leqslant j \leqslant r \\ j \neq k}}
    \frac{\cosh\beta-\cos\alpha_j}{\cos\alpha_k-\cos\alpha_j}\,
    \chi^{so(2,d)}_{\cH}(\beta; \vec \alpha_k)\,.
  \end{split}
  \label{f 1 even AdS}
\end{equation}
The difference between the primary contribution and the full zeta
function, referred to as the secondary contribution, can be computed
to order $z^1$. In \cite{Basile:2018zoy}, it was shown to be absent if
the character of the spectrum is an even function of $\b$. The
higher-spin theories considered in this work fall into this category,
so we will concentrate on the primary contribution in the following
discussions, and omit the subscript 1 in $\zeta_{1,\cH}(z)$\,.

\subsection{Evaluation of the CIRZ Formula}
\label{sec: trick}
The expressions of the CIRZ presented above --- \eqref{GZf} and
\eqref{f even} for even $d$ and \eqref{zeta 1 even AdS} and \eqref{f 1
  even AdS} for odd $d$ --- may look rather implicit compared to the
explicit derivative expansions also presented in
\cite{Basile:2018zoy}, as the $\bm\a$ integrals are left unperformed.
In fact, they prove to be much more useful in actual applications in
this paper, with the help of a few tricks that we shall introduce now.
One of the complications in evaluating the $\a_i$ integral is the
presence of the cyclic permutations over $\a_0,\ldots, \a_r$. Each
permutation has poles of different orders in $\a_i$ and hence
contributes differently. These permutations can be simplified if the
$\a_i$ dependent part of the character can be completely factorized as
\begin{equation}
  \chi^{so(2,d)}_{\cH}(\b,\vec \a)
  =\eta_{\cH}(\b)\,\prod_{i=1}^r\,\xi_{\cH}(\b,\a_i)\,,
  \label{char facto}
\end{equation}
with a function $\xi_{\cH}(\b,\a)$ analytic at $\a=0$.  This is the
case for the scalar and spinor representations and their tensor
products, thereby applicable to the higher-spin gravity theories we
shall consider in the following sections.  With \eqref{char facto},
the $\bm\a$ integral part of the CIRZ formula can be treated for
$d=2r$ as
\begin{eqnarray}
  && \sum_{k=0}^r \oint_C \mu(\bm \alpha)\left(
  \big(\tfrac\beta2\big)^2 + \big(\tfrac{\alpha_k}2\big)^2
  \right)^{z-1}\, \prod_{\substack{0 \leqslant j \leqslant r \\ j \neq
      k}} \left[
    \frac{\cosh\beta-\cos\alpha_j}{\cos\alpha_k-\cos\alpha_j}\,
    \xi_{\cH}(\b,\alpha_j)\right] \nn && = \frac12\, \oint_C \mu(\bm
  \alpha)\,\oint \frac{\dd w}{2\pi\,i} \, \frac{
    \left(\big(\frac\beta2\big)^2 -
    \big(\frac{w}2\big)^2\right)^{z-1}\, \sinh w}{(\cosh\beta-\cosh
    w)\,\xi_{\cH}(\b,i\,w)}\, \prod_{j=0}^r\left[
    \frac{\cosh\beta-\cos\alpha_j}{\cosh
      w-\cos\alpha_j}\,\xi_{\cH}(\b,\alpha_j)\right]\nn &&=
  \frac12\,\oint \frac{\dd w}{2\pi\,i} \, \frac{
    \left(\big(\frac\beta2\big)^2 -
    \big(\frac{w}2\big)^2\right)^{z-1}\, \sinh w}{(\cosh\beta-\cosh
    w)\,\xi_{\cH}(\b,i\,w)}\left[ \frac{\cosh\beta-1}{\cosh
      w-1}\,\xi_{\cH}(\b,0)\right]^{r+1}\,,
  \label{trick_even}
\end{eqnarray}
where the $w$ integration contour  encloses $w=\alpha_i$
anti-clockwise while excluding $\pm \b$.  As a consequence, when the
theory under consideration lies in an odd dimensional AdS background
($d+1=2r+1$) and has the character $\chi^{so(2,d)}_{\cal H}(\b,\vec
\a)$ which can be factorized as \eqref{char facto}, the zeta function
of the theory is
\begin{equation}
  \zeta_{\cal H}(z) = \frac{\ln R}{2\, \Gamma(z)^2}\, \int_0^\infty
  {\dd \beta} \oint \frac{\dd w}{2\pi\,i} \, \frac{
    \left(\big(\frac\beta2\big)^2 -
    \big(\frac{w}2\big)^2\right)^{z-1}\, \sinh
    w\,\eta_{\cH}(\b)}{(\cosh\beta-\cosh w)\,\xi_{\cH}(\b,i\,w)}\left[
    \frac{\cosh\beta-1}{\cosh w-1}\,\xi_{\cH}(\b,0)\right]^{r+1}\,,
    \label{even trick real}
\end{equation}
where the anti-clockwise $w$ contour encloses the origin 
but excludes $\pm \b$.
Using the result of \cite{Basile:2018zoy}, one can further express the
first derivative of the zeta function in terms of contour integrals,
namely
\begin{eqnarray}
  \zeta_{\cal H}'(0) & = & \ln R\, \oint \frac{\dd \beta}{2i\,\pi}\,
  \oint \frac{\dd w}{2\pi\,i}\,\eta_{\cH}(\b)\,
  \big[\xi_{\cH}(\b,0)\big]^{r+1} \times \nn && \qquad \quad \times\,
  \frac{ \sinh w}{\big(\beta^2 - w^2\big)\, (\cosh\beta-\cosh
    w)\,\xi_{\cH}(\b,i\,w)}\left[ \frac{\cosh\beta-1}{\cosh
      w-1}\right]^{r+1}\,.
  \label{even d trick}
\end{eqnarray}
For $d=2r+1$, we introduce an analogous trick:
\begin{eqnarray}
  &&\sum_{k=0}^r \oint_C \frac{\mu(\bm \alpha)\, \big( \cos
    \tfrac{\alpha_k}2
    \big)^{\frac{1-\epsilon}2}}{\cosh\beta-\cos\alpha_k} \prod
  _{\substack{0 \leqslant j \leqslant r \\ j \neq k}}
  \left[\frac{\cosh\beta-\cos\alpha_j}{\cos\alpha_k-\cos\alpha_j}\,
    \xi_{\cH}(\b,\alpha_j)\right] \nn && =\oint_C \mu(\bm \a) \oint
  \frac{\dd w}{2\pi\,i} \frac{\big( \tfrac {w+1}2
    \big)^{\frac{1-\epsilon}4}}{(\cosh\beta-w)^{2}\,\xi_{\cH}(\b,\arccos
    w)} \prod_{j=0}^r
  \left[\frac{\cosh\beta-\cos\alpha_j}{w-\cos\alpha_j}\,
    \xi_{\cH}(\b,\alpha_j)\right]\nn &&= \oint \frac{\dd w}{2\pi\,i}
  \frac{\big( \tfrac {w+1}2
    \big)^{\frac{1-\epsilon}4}}{(\cosh\beta-w)^{2}\,\xi_{\cH}(\b,\arccos
    w)} \left[\frac{\cosh\beta-1}{w-1}\,
    \xi_{\cH}(\b,0)\right]^{r+1}\,,
  \label{trick_odd}
\end{eqnarray}
where the $w$ contour now encircles $w=\cos\a_i$ but excludes $w=\cosh\b$.
For the last equalities in \eqref{trick_even} and \eqref{trick_odd},
the $\alpha_i$ integrals are evaluated independently.  Consequently,
if the theory is in an even dimensional AdS space ($d+1=2r+2$) and has
only bosonic fields, then the primary contribution of the zeta
function is
\begin{eqnarray}
  \zeta_{{\cal H}}(z) & = & \int_0^\infty \frac{\dd
    \beta}{\Gamma(2z)}\, \b^{2z-1}\, \sinh\tfrac\beta2\, \big( \cosh
  \tfrac\beta2 \big)^{\frac{1+\epsilon}2}\, \eta_{\cH}(\b)\, \times
  \nn&& \qquad \times\, \oint \frac{\dd w}{2\pi\,i} \frac{ \big(
    \tfrac {w+1}2 \big)^{\frac{1-\epsilon}4}}{(\cosh\beta-
    w)^{2}\,\xi_{\cH}(\b,\arccos w)} \left[\frac{\cosh\beta-1}{w-1}\,
    \xi_{\cH}(\b,0)\right]^{r+1}\,, \quad
  \label{odd d trick}
\end{eqnarray}
where the anti-clockwise $w$ contour encloses $w=1$ 
but excludes $w=\cosh\b$.
As noted before, the above zeta function is actually the primary
contribution.  The secondary contribution can also be arranged in a
similar manner, but its contribution always vanishes in the
applications we consider in this paper.

\section{Partially Massless Higher-Spin Gravities}
\label{sec:pmhsg}
For $\ell=1$, the type-A$_\ell$ theory coincides with the usual type-A
higher-spin gravity.  For $\ell\ge 2$ the theory involves infinitely
many partially-massless fields besides the massless ones.  In analogy
to the $\ell=1$ case, there are two subclasses: 1) the non-minimal
theory containing fields of all integer spins and 2) the minimal one
containing even spin fields only. Similarly, the type-B$_\ell$ theory
coincides with the usual type-B for $\ell=1$ and also admits a minimal
version.

\subsection{Type-A$_\ell$ field content and characters}
The precise field content of the type-A$_\ell$ higher-spin gravities
is dictated by a generalization of the Flato-Fronsdal theorem
\cite{Bekaert:2013zya}, that is, the tensor product decomposition rule
of the so-called \emph{order-$\ell$} Rac or scalar singleton module,
\begin{equation}
  {\rm Rac}_{\ell}\equiv \mathcal{D}\big( \tfrac{d-2\ell}2; 0 \big)\,,
\end{equation}
where $\mathcal{D}(\D;\Y)$ denotes the irreducible $so(2,d)$ module
with lowest weight $[\Delta;\Y]$. Its character reads
\begin{eqnarray}
  \chi^{so(2,d)}_{\rac_\ell}(\beta, \vec \alpha) \eq
  e^{-\frac{d-2\ell}2\,\b} \, \Pd d (i\beta;\vec
  \alpha)\,(1-e^{-2\,\ell\,\beta}) \nn \eq
  \frac{\sinh(\ell\,\b)}{2^{d-1-r}\, (\sinh\frac\beta2)^{d-2r}}
  \prod_{i=1}^r \frac{1}{\cosh \beta - \cos \alpha_i}\,,
\end{eqnarray}
where
\begin{equation}
  \Pd d (\alpha_0; \vec \alpha) = \frac{e^{-i\,\frac
      d2\,\alpha_0}}{2^{d-r}}\, \prod_{k=1}^r
  \frac{1}{\cos\alpha_0-\cos\alpha_k} \times \left\{
  \begin{aligned}
    &\quad 1 & \qquad [{\rm even}\ d]\\ & \frac i{\sin\frac{\a_0}2} &
    \qquad [{\rm odd}\ d]\\
  \end{aligned}
  \right.\,.
  \label{def_Pd}
\end{equation}
If one considers applying the CIRZ to the Rac$_\ell$ itself --- even
though the module cannot be realized as an AdS field --- we can use
the trick introduced in \hyperref[sec: trick]{Section \ref{sec:
    trick}}, because the character of Rac$_\ell$ can be written as
\eqref{char facto} with
\begin{equation}
  \eta_{\rac_{\ell}}(\b) = \frac{\sinh(\ell\,\b)}{2^{d-1-r}\,
    (\sinh\frac\beta2)^{d-2r}}\,, \qquad \xi_{\rac_{\ell}}(\b,\a) =
  \frac{1}{\cosh \beta - \cos \alpha}\,.
  \label{rac eta xi}
\end{equation}

For the quadratic tensor products of Rac$_\ell$, the only irreps
appearing in the decomposition are $\mathcal{D}\big( s+d-t-1; s\big)$,
which is the spin-$s$ irrep with the lowest energy $s+d-t-1$. Its
character depends on the value of $t$ as
\begin{eqnarray}
  \chi_{[s+d-t-1;s]}^{so(2,d)}(\beta;\vec \alpha) \eq
  e^{-(s+d-t-1)\,\b} \, \Pd d (i\beta;\vec \alpha)\, \times \nn
  &&\quad\times \left\{
  \begin{aligned}
    \chi^{so(d)}_{(s)}(\vec \alpha) \qquad \qquad & \qquad
        [\,t\notin\{1,\ldots,s\}\,] \\ \chi^{so(d)}_{(s)}(\vec \alpha)
        - e^{-\beta\,t}\, \chi^{so(d)}_{(s-t)}(\vec \alpha) & \qquad
        [\,t\in\{1,\ldots,s\}\,]
  \end{aligned}
  \right..
  \label{pm char}
\end{eqnarray}
This representation corresponds to the spin-$s$ (or totally-symmetric
rank-$s$ tensor) field $\varphi_s=\varphi_{\m_1\cdots \m_s}$ in
AdS$_{d+1}$ whose mass depends on the value of $t$.  For
$t=1,\ldots,s$ --- where a submodule structure appears in \eqref{pm
  char} --- the gauge symmetry of the field have the schematic form
\begin{equation}
  \delta_\varepsilon\,\varphi_s = \nabla^t \varepsilon_{s-t}\, ,
\end{equation}
and it is referred to as the spin-$s$ partially-massless field of
depth $t$ \cite{Deser:1983tm, Deser:1983mm, Deser:2001pe,
  Deser:2001xr, Deser:2001us, Higuchi:1986wu}.\footnote{The unfolded
  equation for the free partially-massless fields has been analyzed in
  \cite{Skvortsov:2006at}.  Their cubic interactions have been studied
  in \cite{Boulanger:2012dx, Joung:2012hz, Joung:2012rv}.}  The $t=1$
case corresponds to the massless field and it is the only unitary
irrep.

\paragraph{Non-minimal theory}
The field content of the non-minimal type-A$_\ell$ higher-spin gravity
is given by the Hilbert space $ {\rm A}_{\ell}^{\text{non-min}}$,
isomorphic to the tensor product of two order-$\ell$ Rac modules
\cite{Bekaert:2013zya},
\begin{equation}
 {\rm A}_{\ell}^{\text{non-min}}\cong {{\rm Rac}_{\ell}}^{\otimes 2} \cong
   \bigoplus_{t=1,3,\dots}^{2\ell-1} \bigoplus_{s=0}^\infty
  \mathcal{D}\big( s+d-t-1; s\big)\,.
  \label{FF_Al}
\end{equation}
This spectrum contains the spin $s$ and depth $t$ fields corresponding
to $\mathcal{D}\big(s+d-t-1; s\big)$.  Note that depending on the
value of $t$, it can be either (partially-)massless or not.  The above
decomposition rule can be derived from that of the characters,
\begin{equation}
  \chi^{so(2,d)}_{{\rm A}_{\ell}^{\text{non-min}}}(\b,\vec \a) = \Big(
  \chi^{so(2,d)}_{\rac_\ell}(\beta, \vec \alpha) \Big)^2 =
  \sum_{t=1,3,\dots}^{2\ell-1}\, \sum_{s=0}^\infty
  \chi^{so(2,d)}_{[s+d-t-1;s]}(\beta, \vec \alpha)\,.
  \label{FF_Al_char}
\end{equation}
The above character can be also written as \eqref{char facto} with
\begin{equation}
  \eta_{{\rm A}_{\ell}^{\text{non-min}}}(\b) =
  \frac{\sinh^2(\ell\,\b)}{2^{2(d-1-r)}\,
    (\sinh\frac\beta2)^{2(d-2r)}}, \qquad \xi_{{\rm
      A}_{\ell}^{\text{non-min}}}(\b,\a) = \frac{1}{(\cosh \beta -
    \cos \alpha)^2}\,,
  \label{nm eta xi}
\end{equation}
so we can apply the trick of \hyperref[sec: trick]{Section \ref{sec:
    trick}} for the application of the CIRZ method to this theory.

\paragraph{Minimal theory}
The field content of the minimal type-A$_\ell$ higher-spin gravity is
given by the Hilbert space $ {\rm A}_{\ell}^{\text{min}}$, isomorphic
to the \emph{symmetrized} tensor product of two order-$\ell$ Rac
modules,
\begin{equation}
 {\rm A}_{\ell}^{\text{min}}\cong {{\rm Rac}_{\ell}}^{\odot 2} \cong
 \sum_{t=1,3,\dots}^{2\ell-1}\, \sum_{s=0,2,\dots}^\infty \mathcal D
 \big( s+d-t-1; s \big)\, .
  \label{min_FF_Al}
\end{equation} 
This spectrum is a truncation of the non-minimal one and 
contains only even spin fields.
The above decomposition rule can be derived from that of the characters,
\begin{equation}
  \chi^{so(2,d)}_{{\rm A}_{\ell}^{\text{min}}}(\b,\vec \a)=
  \frac12\,\Big( \chi^{so(2,d)}_{\rac_\ell}(\beta, \vec \alpha)
  \Big)^2 + \frac12\,\chi_{\rac_\ell}^{so(2,d)}(2\beta, 2\vec \alpha)
  = \sum_{t=1,3,\dots}^{2\ell-1}\, \sum_{s=0,2,4,\dots}^\infty
  \chi^{so(2,d)}_{[s+d-t-1;s]}(\beta, \vec \alpha)\,.
  \label{min_FF_Al_char}
\end{equation}
Thanks to the linearity between the zeta function and the character,
we can separately apply the CIRZ to the first and second terms after
the first equality, then sum the results. The first term is nothing
but the half of the non-minimal theory character, so its contribution
to the zeta function is also the half of the non-minimal one. The
second term,
\begin{equation}
  \chi^{so(2,d)}_{{\rm A}_{\ell, {\rm 2nd}}^{\text{min}}}(\b,\vec
  \a)=\frac12\,\chi_{\rac_\ell}^{so(2,d)}(2\beta, 2\vec \alpha)\,,
  \label{min 2nd}
\end{equation}
can also be written as \eqref{char facto} with
\begin{equation}
  \eta_{{\rm A}_{\ell,{\rm 2nd}}^{\text{min}}}(\b) =
  \frac{\sinh(2\,\ell\,\b)}{2^{d-r}\, (\sinh\beta)^{d-2r}}\,, \qquad
  \xi_{{\rm A}_{\ell, {\rm 2nd}}^{\text{min}}}(\b, \a) =
  \frac{1}{\cosh2\beta - \cos2\alpha}\,.
  \label{min 2nd eta xi}
\end{equation}

\subsection{Type-B$_\ell$ field content and characters}
As in the type-A$_{\ell}$ case, the spectrum of the higher-spin theory is
obtained by generalizing the Flato-Fronsdal theorem
\cite{Basile:2014wua} to the tensor product of two order-$\ell$
spin-$\frac12$ singletons\,\footnote{As we shall soon use, in the
  CFT$_d$ this can be realized as an on-shell free conformal (Dirac)
  spinor $\psi$ of conformal weight $\frac{d+1-2\ell}2$, subject to
  the polywave equation $\slashed{\partial}^{2\ell-1} \psi = 0$.}
\begin{equation}
  \di_{\ell} \equiv \mathcal{D}\big( \tfrac{d+1-2\ell}2; \half\big)\,,
\end{equation}
with
\begin{equation}
  \half := \left\{ 
  \begin{aligned}
  (\tfrac12,\dots,\tfrac12,+\tfrac12) \oplus
  (\tfrac12,\dots,\tfrac12,-\tfrac12) \qquad & [d=2r]
  \\ (\tfrac12,\dots,\tfrac12)\, \qquad \qquad \qquad & [d=2r+1]
  \end{aligned}
  \right.\,.
\end{equation}
In other words, we will consider the parity-invariant spin-$\tfrac12$
singleton, the character of which reads
\begin{eqnarray}\label{eq:char_di_ell}
  \chi^{so(2,d)}_{\di_\ell}(\beta, \vec \alpha) & = &
  e^{-\beta(\frac{d+1-2\ell}2)} (1-e^{-(2\ell-1)\beta}) \,
  \chi^{so(d)}_{\boldsymbol{\frac12}}(\vec \alpha)\, \Pd d
  (i\beta;\vec \alpha) \nn & = & \frac{\sinh(\beta\,
    \tfrac{2\ell-1}2)}{2^{d-2r-1}\, (\sinh\tfrac\beta2)^{d-2r}}
  \prod_{i=1}^r \frac{\cos\tfrac{\alpha_i}2}{\cosh \beta - \cos
    \alpha_i}\,.
\end{eqnarray}
This character can be also written as \eqref{char facto} with
\begin{equation}
  \eta_{\di_{\ell}}(\b) =
  \frac{\sinh(\b\,\frac{2\ell-1}2)}{2^{d-2r-1}\,
    (\sinh\frac\beta2)^{d-2r}}, \qquad 
    \xi_{\di_{\ell}}(\b,\a) = \frac{\cos\frac\a2}{\cosh \beta -
    \cos \alpha}\,.
  \label{di eta xi}
\end{equation}
The tensor product of two $\di_\ell$ decomposes into a direct sum of
irreps $\mathcal{D}\big( s+d-t-1; s,1^m\big)$. Here $(s,1^m)$ with $s
\ge 1$ and $m=0,\dots,r-1$ is a shorthand notation used to denote the
$so(d)$ weight
\begin{equation}
  (s,1^m) := ( \ s,\,\,\underbrace{1,\dots,1}_{m\, {\rm terms}},
  \underbrace{0,\dots,0}_{r-1-m\, {\rm terms}})\,.
\end{equation}
Fields of spin $(s,1^m)$ with $m \ge 1$ are the simplest types of
mixed-symmetry fields. Note that this contrasts with the
type-A$_{\ell}$ theories, whose spectrum do not contain any mixed-symmetry
representations. The character of such fields is given by
\begin{equation}
  \begin{split}
    \chi_{[s+d-t-1;s,1^m]}^{so(2,d)}(\beta;\vec \alpha) =
    &e^{-\beta(s+d-t-1)} \, \Pd d (i\beta;\vec \alpha)\, \\&\times
    \left\{
    \begin{aligned}
      \chi^{so(d)}_{(s,1^m)}(\vec \alpha) \qquad \qquad & \quad
          [\,t\notin\,\{1,\dots,s\}] \\ \chi^{so(d)}_{(s,1^m)}(\vec
          \alpha) - e^{-\beta t}\, \chi^{so(d)}_{(s-t,1^m)}(\vec
          \alpha) & \quad [\,t\in\,\{1,\dots,s\}]
    \end{aligned}
    \right..
  \end{split}
\end{equation}
As in the type-A$_\ell$ case, these irreps are unitary only for the
$t=1$ case.  The latter corresponds to mixed-symmetry massless fields
whose study was initiated by Metsaev in \cite{Metsaev:1995re,
  Metsaev:1997nj, Metsaev:1998xg} (see also \cite{Burdik:2000kj,
  Burdik:2001hj, Alkalaev:2003qv, Bekaert:2003az, Bekaert:2003zq,
  Alkalaev:2008gi, Skvortsov:2009nv, Skvortsov:2009zu,
  Campoleoni:2009je, Alkalaev:2009vm, Alkalaev:2011zv,
  Campoleoni:2012th} and references therein). When $1<t \le s$, the
irreps correspond to mixed-symmetry partially-massless depth-$t$
AdS$_{d+1}$ fields \cite{Boulanger:2008up,Boulanger:2008kw} which are
dual to partially-conserved mixed-symmetry CFT$_d$ currents
\cite{Alkalaev:2012rg, Alkalaev:2012ic,Chekmenev:2015kzf}. Finally,
when $s<t$, these modules correspond to massive AdS fields of minimal
energy $s+d-t-1$.\\

The spectrum of the non-minimal type-B$_\ell$ higher-spin theory is given by
the tensor product of two spin-$\frac12$ singleton of order-$\ell$
\cite{Basile:2014wua}:
\begin{eqnarray}
  \mathcal{D}\big( \tfrac{d+1-2\ell}2; \half \big)^{\otimes 2} & \cong
  & \bigoplus_{t=-2(\ell-1)}^{2(\ell-1)} \mathcal{D}\big( d-t-1; 0
  \big) \oplus \bigoplus_{t=1}^{2\ell-1}\bigoplus_{s=1}^\infty
  \bigoplus_{m=0}^{r-1} \mathcal{D}\big( s+d-t-1; s,1^m \big)
  \nonumber \\ && \qquad \qquad \oplus
  \bigoplus_{t=1}^{2(\ell-1)}\bigoplus_{s=1}^\infty
  \bigoplus_{m=0}^{r-1} \mathcal{D}\big( s+d-t-1; s,1^m \big)\,.
  \label{FF_Bl}
\end{eqnarray}
The spectrum of the minimal type-B$_{\ell}$ theory is given by the
antisymmetric tensor product of two spin-$\frac12$ singleton of
order-$\ell$.  Its explicit content is however more complicated and
the closed form expression is relegated to
\hyperref[app:minBl]{Appendix \ref{app:minBl}}.  For the purposes of
our computations, it is sufficient to specify their characters:
\begin{equation}\label{char_di_nonmin}
  \chi^{so(2,d)}_{\text{B}_{\ell}^{\text{non-min}}}(\beta,\vec\alpha)
  = \chi^{so(2,d)}_{\di_\ell^{\otimes 2}}(\beta, \vec\alpha) =
  \Big(\chi^{so(2,d)}_{\di_\ell}(\beta, \vec\alpha) \Big)^2\,,
\end{equation}
\begin{equation}
  \chi^{so(2,d)}_{\text{B}_{\ell}^{\text{min}}}(\beta,\vec\alpha) =
  \chi^{so(2,d)}_{\di_{\ell}^{\wedge 2}}(\beta, \vec\alpha) = \tfrac12
  \Big(\chi^{so(2,d)}_{\di_{\ell}}(\beta, \vec\alpha)\Big)^2 -\tfrac12
  \chi^{so(2,d)}_{\di_{\ell}}(2\beta,2\vec\alpha) \,.
  \label{chi-Bl-min}
\end{equation}
From \eqref{eq:char_di_ell} and \eqref{char_di_nonmin}, we may read
off the functions
\begin{equation}
  \eta_{\rm B_{\ell}^{\text{non-min}}}(\beta) = \frac{\sinh^2(
    \tfrac{2\ell-1}2\,\beta)}{2^{2(d-2r-1)}\,
    (\sinh\tfrac\beta2)^{2(d-2r)}}\,, \quad \xi_{\rm
    B_{\ell}^{\text{non-min}}}(\beta, \alpha) =
  \frac{\cos^2\tfrac\alpha2}{(\cosh \beta - \cos \alpha)^2}\,,
  \label{etaxiBl}
\end{equation}
for the non-minimal type-B$_\ell$ theory. Similarly to the case of the
minimal type-A$_\ell$, the zeta function of the minimal type-B$_\ell$
is given by two terms corresponding to the last expression in
\eqref{chi-Bl-min}. The first one is simply half of the zeta function
of the non-minimal theory, and hence it will prove useful to introduce
the quantities
\begin{equation}
  \eta_{\rm B_{\ell, 2nd}^{\rm min}}(\beta) = -\frac{\sinh\big(
    (2\ell-1)\,\beta\big)}{2^{d-2r}\, (\sinh\beta)^{d-2r}}\, , \qquad
  \xi_{\rm B_{\ell, 2nd}^{min}}(\beta, \alpha) =
  \frac{\cos\alpha}{\cosh 2\beta - \cos 2\alpha}\,,
  \label{etaxiBlmin}
\end{equation}
for the contribution of the second term.

\section{Type-A$_\ell$ higher-spin gravities}
\label{sec:application}
Now, we are ready to apply the CIRZ method to the type-A$_\ell$
higher-spin gravity theories. Using the tricks introduced in
\hyperref[sec: trick]{Section \ref{sec: trick}}, the zeta function can
be simplified to the form of \eqref{even d trick} and \eqref{odd d
  trick}. In the following, we divide the task into two parts: first,
the case of even $d$, and then that of odd $d$.

\subsection{AdS$_{2r+1}$}\label{sec:application d eq 2r}
In odd dimensional AdS, we can use the expression \eqref{even d trick}
for the zeta function. As discussed in \hyperref[sec: trick]{Section
  \ref{sec: trick}}, the zeta function manifestly vanishes at $z=0$
due to the presence of $1/\Gamma(z)^2$, which is consistent with the
well-known fact that odd dimensional theories have no logarithmic
divergences. On the other hand, the derivative of the zeta function at
$z=0$ is given by a contour integral in $\beta$ around the origin. In
the following, we shall directly focus on the derivative of the zeta
function.

\paragraph{Non-minimal  theory}
\label{zeta_Al_even}
Inserting the functions \eqref{nm eta xi} into \eqref{even d trick},
we obtain
\begin{equation}
  \zeta_{{\rm A}_{\ell}^{\text{non-min}}}'(0) = \frac{\ln
    R}{2^{2r-2}}\, \oint\frac{\dd \beta}{2\pi\,i}\, \oint \frac{\dd
    w}{2\pi\,i}\,\frac1{\b^2-w^2}\, \frac{\sinh^2(\ell\,\b)\,\sinh
    w\,(\cosh\beta-\cosh w)}{(\cosh\b-1)^{r+1}\,(\cosh w-1)^{r+1}}\,.
  \label{A nm ct}
\end{equation}
Due to the fact that the integrand is an even function of $\beta$, the
contour integral trivially vanishes, and hence we conclude that
\begin{equation}\label{zeta al nonmin 2r+2}
  \zeta'_{{\rm A}_{\ell}^{\text{non-min}}}(0)=0\,.
\end{equation}
Let us remark one subtlety in transforming the real $\b$ integral
\eqref{even d trick} to the contour one \eqref{A nm ct}: the integrand
behaves as $e^{-\beta(\frac d2-2\ell)}$ when $\beta\rightarrow\infty$,
and therefore the integral over real $\b$ would diverge unless
$\ell<\tfrac d4$.\,\footnote{This bound is somewhat surprising, as it
  is in fact more constraining than the bound $\ell<\tfrac d2$ found
  on the CFT side for the convergence of the zeta function (see the
  discussion below \eqref{zeta_rac} in the next subsection).}  This
divergence can be traced back to the Camporesi-Higuchi formula
\cite{Camporesi1994} where the $u$-integral diverges as $u\to 0$ for
$\bar\Delta=\D-\tfrac{d}{2}=0$.  In the type-A$_\ell$ higher-spin
gravity theories, the spin-$s$ and depth-$t$ fields have $\Delta =
s+d-t-1$ with $t=1,3,\dots,2\ell-1$.  Therefore, this type of
singularity arises unless $\Delta >\tfrac d2$ for all spin $s$, which
is equivalent to $\ell<\tfrac d4$.  The remedy we adopt for this
divergence, both on the AdS and the CFT side, is to work with a value
of $\ell$ such that the $\beta$ integrals converge, and then
analytically continue the obtained results to arbitrary values of
$\ell$.  This regularization is consistent with the one used in
\cite{Giombi:2013fka, Giombi:2014iua} for the $\ell=1$ case.

\paragraph{Minimal theory}
\label{zeta_Al_min_even}
The zeta function of the minimal type-A$_\ell$ higher-spin gravity has
two parts. The first part is equal to the half of the zeta function of
the minimal theory. Since we have just shown that the minimal theory
gives a vanishing zeta function, up to the physically relevant order
of ${\cal O}(z^2)$\,, we focus on the second part with the character
\eqref{min 2nd}. Substituting \eqref{min 2nd eta xi} into \eqref{even
  d trick}, we arrive at
\begin{eqnarray}\label{minzeta 1}
  \zeta'_{{\rm A}_{\ell}^{\text{min}}}(0) & = & \frac{\ln R}{2^{2r}}\,
  \oint \frac{\dd \b}{2\pi\,i}\, \oint \frac{\dd w}{2\pi\,i}\,
  \frac1{\b^2 - w^2}\, \frac{\sinh(2\,\ell\,\b)\,\sinh
    w\,(\cosh\beta+\cosh w)}{(\cosh\b+1)^{r+1}\,(\cosh w-1)^{r+1}}\,.
\end{eqnarray}
The contour integral with respect to $w$ contains an order $2r+2$ pole
at $w=0$, and hence is rather cumbersome to evaluate for an arbitrary
$r$. Instead, we can first perform the $\b$ contour which contains
only two simple poles at $\beta=\pm w$. Then, we end up with the $w$
integral
\begin{equation}
  \zeta'_{{\rm A}_{\ell}^{\text{min}}}(0) = \frac{\ln R}{2^{2r-1}}\,
  \oint \frac{\dd w}{2\pi\,i\,w}\, \frac{\sinh(2\,\ell\,w)\,\cosh
    w}{(\sinh w)^{2r+1}}\,.
  \label{min zeta}
\end{equation}
The evaluation of the above gives a polynomial in $\ell$ of order
$2r+1$.
Later, we will show that the same contour
integral appears from the CFT zeta function.
 
\paragraph{Order-$\ell$ Rac module}
It is interesting to compare the result \eqref{min zeta} of the
minimal theory with that of the order-$\ell$ Rac module Rac$_\ell$.
Since the CIRZ formula is defined for any $so(2,d)$ character, one can
consider the AdS zeta function of Rac$_\ell$ by treating the module as
if it can be realized as an AdS field.  Inserting \eqref{rac eta xi}
into \eqref{even d trick}, the derivative of the zeta function for
Rac$_\ell$ is given by
\begin{equation}
  \zeta'_{{\rm Rac}_{\ell}}(0)= \frac{\ln R}{2^{r-1}}\, \oint
  \frac{\dd \b}{2\pi\,i}\, \oint \frac{\dd w}{2\pi\,i}\, \frac1{\b^2 -
    w^2}\, \frac{\sinh(\ell\,\b)\,\sinh w}{(\cosh w-1)^{r+1}}\,.
\end{equation}
Again evaluating the $\b$ integral first, we obtain
\begin{equation}\label{zetaracelladsodd}
  \zeta'_{{\rm Rac}_{\ell}}(0)= \frac{\ln R}{2^{2r-1}}\, \oint
  \frac{\dd w}{2\pi\,i\,w}\, \frac{\sinh(\ell\,w)\,\cosh \frac
    w2}{(\sinh\frac w2)^{2r+1}}\,,
\end{equation}
which coincides with \eqref{min zeta} upon the rescaling of the
variable $w$.

\subsection{AdS$_{2r+2}$}
We now turn to the case of even dimensional AdS where we will use the
expression \eqref{odd d trick} for the zeta function. We emphasize
again that in general there is an additional contribution to the zeta
function, which identically vanishes to $\mathcal{O}(z^2)$ for the
Type-A$_\ell$ theories as their character is an even function of $\b$.

\paragraph{Non-minimal theory}
Inserting the functions \eqref{nm eta xi} into \eqref{odd d trick}, we
obtain
\begin{equation}
  \zeta_{{\rm
      A}_\ell^{\textrm{non-min}}}(z)=\frac1{2^{2r+1}}\int_0^\infty
  \frac{\dd \beta}{\Gamma(2z)}\, \oint \frac{\dd w}{2\pi\,i}
  \frac{\b^{2z-1}\,\sinh\b\,\sinh^2(\ell\b)}{
    \sinh^2\frac\b2\,(\cosh\beta-1)^{r+1}(w-1)^{r+1}}\,.
\end{equation}
It is simple to check that the $w$ integral vanishes. Consequently,
the (primary contribution) of the zeta function simply vanishes: 
\be
\zeta_{{\rm
    A}_\ell^{\textrm{non-min}}}(z)=0\,.
\ee

\paragraph{Minimal theory}

The character for the minimal theory is given by
\eqref{min_FF_Al_char}. From the analysis for the non-minimal theory
we saw already that the contribution to the zeta function vanishes up
to $\mathcal{O}\left(z^2\right)$ terms.  We only need to apply the
CIRZ to the second term \eqref{min 2nd} to obtain the zeta function
for the minimal theory.  Applying \eqref{min 2nd eta xi} to \eqref{odd
  d trick} we obtain
\begin{equation}
  \zeta_{{\rm A}_{\ell,{\rm 2nd}}^{\textrm{min}}}(z) =
  \frac1{2^{2r+2}}\int_0^\infty \frac{\dd \beta}{\Gamma(2z)}\, \oint
  \frac{\dd w}{2\pi\,i} \frac{\b^{2z-1}\,\sinh(2\,\ell\,\b)\,
    (\cosh\b+w)}{(\cosh\beta- w)\,(\cosh\beta+1)^{r+1}\,
    (w-1)^{r+1}}\,.
\end{equation}
We can evaluate the $w$ integral (for $r\geqslant 1$) as
\begin{equation}
  \oint \frac{\dd w}{2\pi\,i}\,\frac{\cosh\b+w}{\cosh\beta-w}\,
  \frac1{(w-1)^{r+1}} = \frac{2\,\cosh\b}{(\cosh\b-1)^{r+1}}\,.
\end{equation}
Hence, the zeta function for the non-minimal theory reduces to
\begin{equation}
  \zeta_{{\rm A}_{\ell}^{\textrm{min}}}(z) =
  \frac1{2^{2r+1}}\int_0^\infty \frac{\dd \beta}{\Gamma(2z)}\,
  \frac{\b^{2z-1}\,\sinh(2\,\ell\,\b)\,\cosh\b}{(\sinh\b)^{2r+2}}\,.
  \label{min int}
\end{equation}
Therefore, one can easily conclude that the zeta functions vanish at
$z=0$,
\begin{equation}
  \zeta_{{\rm A}_{\ell}^{\textrm{min}}}(0) = 0\,,
\end{equation}
from the fact that the integrand is an even function of $\b$ for
$z=0$. Let us postpone the extraction of the derivative of the zeta
function from the integral \eqref{min int} for a while, because the
integral \eqref{min int} itself can be matched to the CFT side.

\paragraph{Order-$\ell$ Rac module}
As a prelude to the explicit computation on the CFT, we also follow
the AdS$_{2r+1}$ analysis and formally treat the Rac$_\ell$ module as
a field in AdS$_{2r+2}$ and compute its one-loop determinant and hence
free energy. Substituting \eqref{rac eta xi} into \eqref{odd d trick},
we obtain
\begin{equation}
  \zeta_{{\rm Rac}_{\ell}}(z) = \frac1{2^{r+1}}\int_0^\infty \frac{\dd
    \beta}{\Gamma(2z)}\, \oint \frac{\dd w}{2\pi\,i}
  \frac{\b^{2z-1}\,\sinh(\ell\,\b)\,\sinh\b}{\sinh\frac\b2\,(\cosh\beta-
    w)\,(w-1)^{r+1}}\,.
\end{equation}
After the $w$ integral, it becomes
\begin{equation}\label{zetaracelladseven}
  \zeta_{{\rm Rac}_{\ell}}(z) = \frac1{2^{2r+1}}\int_0^\infty
  \frac{\dd \beta}{\Gamma(2z)}\, \frac{\b^{2z-1}\, \sinh(\ell\,\b)\,
    \cosh\frac\b2}{(\sinh\frac\b2)^{2r+2}}\,,
\end{equation}
and hence can be related to the minimal model zeta function as
\begin{equation}
  \zeta_{{\rm Rac}_{\ell}}(z)=4^z\, \zeta_{{\rm
      A}_{\ell}^{\textrm{min}}}(z)\,.
  \label{rel_Amin_rac_odd}
\end{equation}
Since these zeta functions vanish at $z=0$, the above implies that
their first derivatives at $z=0$ coincide with each other.

\subsection{CFT$_d$}
In the previous subsection, we have shown that for any $d$, the zeta
function of the non-minimal type-A$_\ell$ higher-spin gravity in
AdS$_{d+1}$ vanishes up to $\cO(z^2)$, and hence so does its one-loop
free energy.  This confirms the AdS/CFT duality that we reviewed in
the beginning of the current section. We obtained an integral
expression for the zeta function of the non-minimal theory, which
coincides with that of the order-$\ell$ Rac module. Since it is not
obvious whether or not the AdS$_{d+1}$ free energy for $\rac_\ell$
would be the same as the $S^d$ free energy of the order-$\ell$ free
scalar, we calculate hereafter the free energy of the latter. Notice
that this calculation has been carried out
previously\,\footnote{See also \cite{Dowker:2017qkx} for
    computations of the R\'enyi entropies and central charges of the
    higher-order scalar and spinor singletons, as well as
    \cite{Dowker:2016fqv} for computations of their Casimir energy.}
in \cite{Brust:2016gjy} for odd dimensions up to $d=13$ and $\ell=1,2,
3$ (whereas previous computations for the unitary conformal scalar
field, i.e. $\ell=1$, can be found in e.g. \cite{Gubser:2002vv,
  Klebanov:2011gs, Diaz:2007an}). We will start by revisiting the
computation of the $\rac_\ell$ zeta function, so as to express it in
term of the character of the order-$\ell$ scalar.\\

The order-$\ell$ scalar singleton in $d$-dimensions can be realized as
a free conformal scalar field defined by a $2\ell$-derivative
action. In flat space, the action reads
\begin{equation}
  S_{\rac_\ell}[\phi] = \int \dd^d x\, \phi\, \Box^\ell\, \phi\,,
\end{equation}
where we can see that the conformal weight of $\phi$ is
$\frac{d-2\ell}2$\,. When the background is a $d$-dimensional sphere,
the action becomes\,\footnote{For generic Einstein manifolds, the
  action requires specific conformal couplings, which have been
  determined in \cite{Juhl2009, Juhl2013, Fefferman2012,
    Beccaria:2015vaa}.}
\begin{equation}
  S_{\rac_\ell}[\phi] = \int \dd^d x\,\sqrt{g}\, \phi\,
  \prod_{k=1}^\ell \Big( \nabla_{S^d}^2 - (\tfrac
  d2-k)(\tfrac{d-2}2+k) \Big)\, \phi\,.
 \label{ell S}
\end{equation}
where $\nabla^2_{S^d}$ is the Laplace-Beltrami operator on the
$d$-dimensional unit sphere. The eigenvalues of $\nabla^2_{S^d}$
acting on scalar fields on $S^d$ are $-n(n+d-1)$ with $n\in\N$, and
hence the eigenvalues $\lambda_n^{(\ell)}$ of the order-$2\ell$ wave
operator in the action are the product
\begin{equation}
  \lambda^{(\ell)}_n = \prod_{k=0}^{2\ell-1} \lambda_{n,k}^{(\ell)}\,
  , \qquad \lambda_{n,k}^{(\ell)} := \tfrac{d-2\ell}2+n+k\,.
  \label{eigenvalue lambda}
\end{equation}
The degeneracies $d_n$ for a given $n$ is independent of $\ell$ and
given by
\begin{equation}
  d_n = \frac{(d+2n-1)\, (d+n-2)!}{n!\,(d-1)!}  \equiv
  \dim^{so(d+1)}_{(n)}\,.
\end{equation}
This information implies that the free energy of \eqref{ell S} is a
divergent series:
\begin{equation}
  F_{\rac_\ell} = \frac12\, \sum_{n=0}^\infty d_n\, \ln
  \lambda^{(\ell)}_n = \frac12 \sum_{k=0}^{2\ell-1} \sum_{n=0}^\infty
  \dim_{(n)}^{so(d+1)}\, \ln\lambda^{(\ell)}_{n,k}\,.
  \label{free_energy_rac}
\end{equation}
We can regularize the series through the zeta function method as in
the bulk theory. Hence, we will consider the zeta
function\,\footnote{Remark also that the replacement of $\ln A$ by
  $A^{-z}$ in the zeta function regularization can be done at various
  stages. For instance, $\ln(A_1\,A_2\,\cdots)$ can be directly
  replaced by $(A_1\,A_2\,\cdots)^{-z}$ or first decomposed into
  $\ln(A_1)+\ln(A_2)+\cdots$ then replaced by
  $(A_1)^{-z}+(A_2)^{-z}+\cdots$. Different choices sometimes give
  different results, and this phenomenon is referred to as
  ``multiplicative anomaly''. The choice we make is the replacement
  after full decomposition of the logarithms. See
  e.g. \cite{Brust:2016xif} and references therein.}
\begin{equation}
  \zeta^{\sst(d)}_{\rac_\ell}(z) = \sum_{k=0}^{2\ell-1}
  \sum_{n=0}^\infty \dim_{(n)}^{so(d+1)}
  \left(\lambda^{(\ell)}_{n,k}\right)^{-z}\,.
  \label{zeta_rac_sum}
\end{equation}
Following \eqref{free energy}, the free energy can be related to the
zeta function as
\begin{equation}\label{fracelld}
 F_{\rac_\ell} = -\frac12\,{\zeta^{\sst (d)}_{\rac_\ell}}(0)\ln 
 \left(R\,\Lambda_{\rm\sst UV}\right)
  -\frac12\,{\zeta^{\sst (d)}_{\rac_\ell}}'(0)\,.
\end{equation}
Here $\Lambda_{\rm\sst UV}$ is a UV cutoff which is multiplied to the
radius $R$ of $S^d$ for dimensional reasons. It is suppressed in the
expressions that follow. We have used the notation $\zeta^{\sst
  (d)}_{\rac_\ell}(z)$ to stress that the zeta function is computed on
$S^d$. This is a priori different from the AdS zeta function
$\zeta_{\rac_\ell}(z)$.

Now, we shall re-express the zeta function \eqref{zeta_rac_sum} as a
Mellin integral form. First, we transform it into
\begin{equation}
  \zeta^{\sst (d)}_{\rac_\ell}(z) = \int_0^\infty \frac{\dd
    \beta}{\Gamma(z)}\, \beta^{z-1}\, \sum_{k=0}^{2\ell-1}
  \sum_{n=0}^\infty \dim_{(n)}^{so(d+1)}
  e^{-(\frac{d-2\ell}2+n+k)\,\b}\,,
  \label{zeta bd}
\end{equation}
then perform the summation over $k$ and $n$ using the identity,
\begin{equation}
  \sum_{n=0}^\infty \dim_{(n)}^{so(d+1)} \,e^{-n\,\b}=(1-e^{-2\,\b})\,
  \Pd{d+1}(i\b;0) = (1+e^{-\b})\,\Pd{d}(i\b;0)\,.
\end{equation}
Finally, the zeta function can be written as
\begin{eqnarray}
  \zeta^{\sst (d)}_{\rac_\ell}(z) \eq \int_0^\infty \frac{\dd
    \beta}{\Gamma(z)}\, \beta^{z-1}\, e^{-\beta (\frac{d-2\ell}2)}\,
  \frac{1-e^{-2\beta\ell}}{1-e^{-\beta}}\, (1+e^{-\beta})\,\Pd d
  (i\beta;0) \nn \eq \int_0^\infty \frac{\dd \beta}{\Gamma(z)}\,
  \beta^{z-1}\, \coth\tfrac\beta2\, \chi_{{\rm
      Rac}_\ell}^{so(2,d)}(\beta,\vec 0),
  \label{zeta_rac}
\end{eqnarray}
or more explicitly
\begin{equation}
  \zeta^{\sst (d)}_{\rac_\ell}(z) = \frac1{2^{d-1}} \int_0^\infty
  \frac{\dd \beta}{\Gamma(z)}\, \frac{\beta^{z-1}\, \sinh(\ell\,\b)\,
    \cosh\frac\b2}{(\sinh\frac\b2)^{d+1}}\,.
\end{equation}
The above integral behaves as $e^{-\frac{d-2\ell}2\,\b}$
asymptotically for large $\b$, and hence it is convergent as long as
$\ell<\tfrac{d}{2}$, or equivalently, the conformal weight of
$\rac_\ell$ is positive. One can still consider the case with the
negative conformal weight as an analytic continuation in $\ell$ or
$d$. Furthermore, we can explicitly evaluate \eqref{zeta_rac} in terms
of the Lerch transcendent $\Phi(p,z,a)$:
\begin{eqnarray}
  \zeta^{\sst (d)}_{\rac_\ell}(z) \eq \frac{1}{d!}\,
  \Big(\frac\partial{\partial p}\Big)^d\, \Big[\,
    \Phi(p,z,-\tfrac{d+2\,\ell}2) - \Phi(p,z,-\tfrac{d-2\,\ell}2) \nn
    && \hspace{60pt} +\, \Phi(p,z,1-\tfrac{d+2\,\ell}2) -
    \Phi(p,z,1-\tfrac{d-2\,\ell}2) \Big]_{p=1}\, .
  \label{zeta_rac_lerch}
\end{eqnarray}
Note that the above expression holds for both even and odd $d$.  Since
$\partial_p^n\,\Phi(p,z,a)|_{p=1}$ reduces to a sum of Hurwitz-zeta
function $\zeta(z-k,a)$ with $k=0,\dots,n$, the right hand side of the
equality in \eqref{zeta_rac_lerch} can be expressed as a linear
combination of Hurwitz zeta functions. Eventually, we can take the
first derivative in $z$. For the further analysis, we need to
distinguish again the case in even dimensions from that in odd
dimensions.

\subsubsection{CFT$_{2r}$}
\label{boundaryeven}
In even boundary dimensions, i.e. when $d=2r$, we can obtain
${\zeta^{\sst (2r)}_{\rac_\ell}}(0)$ from \eqref{zeta_rac} as the
contour integral
\begin{equation}
  \zeta^{\sst (2r)}_{\rac_\ell}(0) = \frac1{2^{2r-1}} \oint \frac{\dd
    \beta}{2\pi\,i\,\b}\, \frac{\sinh(\ell\,\b)\,
    \cosh\frac\b2}{(\sinh\frac\b2)^{2r+1}} \,.
  \label{zeta d even}
\end{equation}
By comparing it with $\zeta'_{{\rm A}_{\ell}^{\text{min}}}(0)$ in
\eqref{min zeta}, we find that the two contour integral expressions
coincide up to $\ln R$\,:
\begin{equation}
  \zeta'_{{\rm A}_{\ell}^{\text{min}}}(0) = \ln R\,\zeta^{\sst
    (2r)}_{\rac_\ell}(0) =-2\,\ln R\,a_{\rac_\ell}\,,
\end{equation}
where ${\zeta^{\sst (2r)}_{\rac_\ell}}(0)$ is related to the Weyl
anomaly $a$ coefficient by $a_{\rac_\ell} = -\frac12\,
\zeta_{\rac_\ell}^{\sst (2r)}(0)$.  We do not need the explicit values
of the integrals \eqref{min zeta} and \eqref{zeta d even}, but they
can be evaluated readily by computing the residue of the integrand in
\eqref{zeta d even}
\begin{equation}
  \zeta^{\sst (2r)}_{\rac_\ell}(0) = \frac{1}{(2r)!}\,
  \Big(\frac{\dd}{\dd \beta}\Big)^{2r}\, \left[
    \frac{\beta^{2r}}{2^{2r-1}}\, \frac{\sinh(\ell\,\beta)\,
      \cosh\frac\b2}{(\sinh\frac \b2)^{2r+1} } \right]\Bigg|_{\beta=0} \, .
\end{equation}
The corresponding $a$-anomaly coefficients in a few low-$d$ cases
are summarized in \hyperref[table]{Table \ref{table bos}}.
\begin{table}[h!]
  \centering
  \begin{center}
    \begin{tabular}{c|c}
      $d$ & $a_{\rac_\ell}$ \\[5pt] 
      \hline\hline
      $2$ & $-\frac{1}3\ell^3$ \\[5pt]
      \hline 
      $4$ & $\frac{1}{180}\ell^3(5-3\ell^2)$ \\[5pt]
      \hline     
      $6$ & $-\frac{1}{7560}\ell^3(28-21\ell^2+3\ell^4)$ \\[5pt]
      \hline      
      $8$ & $\frac{1}{907200}\ell^3(540 - 441\ell^2 + 90\ell^4 -
      5\ell^6)$ \\[5pt]
      \hline
      $10$ & $-\frac{1}{59875200}\ell^3(6336 - 5412 \ell^2+ 1287
      \ell^4 - 110 \ell^6 + 3 \ell^8)$
    \end{tabular}
  \end{center}
  \caption{Summary of $a$-anomaly coefficients for the order-$\ell$
    real scalar in low dimensions.}
  \label{table bos}
\end{table}

\subsubsection{CFT$_{2r+1}$}
\label{boundaryodd}
In odd boundary dimensions, i.e. when $d=2r+1$, we first find that the
zeta function of the order-$\ell$ scalar on $S^d$ is related to the
primary contribution of the zeta function of the type-A$_\ell$ minimal
higher-spin gravity as
\begin{equation}
  \zeta_{{\rm A}_{\ell}^{\text{min}}}(z)= 2^{-2z-1}\, \zeta^{\sst
    (2r)}_{\rac_\ell}(2z)\,.
\end{equation}
These zeta functions vanish at $z=0$ because the integrand of the
contour integral is an even function:
\begin{equation}
  \zeta^{\sst (2r+1)}_{\rac_\ell}(0) = 0\,.
\end{equation}
This is of course a general property of even dimensional theories.
Moving to the derivative of the zeta function, we find
\begin{equation}
  \zeta\,'_{{\rm A}_{\ell}^{\text{min}}}(0) = \zeta^{\sst
    (2r+1)}_{\rac_\ell}{}'(0)\,.
\end{equation}

As already noticed in \cite{Brust:2016gjy}, the free energy of the
minimal type-A$_\ell$ higher spin gravity or the order-$\ell$ scalar
CFT develops an imaginary part for $\ell > \tfrac d2$.  For instance,
in $d=3$ dimensions, we find
\begin{equation}
    \zeta\,'_{{\rm A}_{\ell}^{\text{min}}}(0) = \zeta^{\sst
      (3)}_{\rac_\ell}{}'(0)= -\frac{1}{8} \Big(
    \tfrac23\,\ell\,(4\ell^2-1) \ln 2 - \frac{3\ell}{\pi^2} \zeta(3)
    \Big) -i\,\frac{\pi}{12}\,\ell^2(\ell^2-1)\,.
\end{equation}
From the CFT point of view, the imaginary number arises from the terms
in the summand with negative eigenvalue $\l_{n,k}^{\sst (\ell)}$
\eqref{eigenvalue lambda} in the free energy \eqref{free_energy_rac}
or equivalently, in the zeta function \eqref{zeta_rac_sum}.  Clearly,
this happens when $\tfrac{d-2\ell}2=r-\ell+\frac12<0$.  By introducing
$m=\ell-r-1$, we can write the imaginary part as the finite sum
\begin{equation}
  i\,{\rm Im}(F_{\rac_\ell}) = i\,\frac{\pi}2\, \sum_{k=0}^{m}
  \sum_{n=0}^{m-k} \dim_{(n)}^{so(2r+2)}\,.
\end{equation}
Performing the summation, we obtain
\begin{eqnarray}
  i\,{\rm Im}( F_{\rac_\ell} ) =
  i\,\frac{\pi}{(2r+2)!}\,\prod_{n=0}^r(\ell^2-n^2)\,.
\end{eqnarray}
Notice that the imaginary part vanishes for $1 \leqslant \ell
\leqslant r$, which is consistent with the previous discussion.  From
the AdS point of view, the imaginary part appears from the finite
subset of the spectrum with negative $\D$. For such fields, the
$\beta$ integral in the zeta function is not convergent in the large
$\b$ region.

\subsection{``Generalized'' free energy from the AdS perspective}
\label{sec: gen free}
An expression for the ``generalized'' sphere free energy $\tilde F$
was (defined and) proposed in \cite{Giombi:2014xxa} (see also
\cite{Dowker:2017hpm} for a generalization). It interpolates between
$(-1)^{d/2} \tfrac\pi2$ times the Weyl anomaly coefficient in even
dimensions and $(-1)^{(d-1)/2}$ times the free energy in odd
dimensions. For the unitary conformal scalar, i.e. whose conformal
weight is $\tfrac{d-2}2$, this quantity is given by
\begin{equation}
  \tilde F = \frac{1}{\Gamma(d+1)}\, \int_0^1 \dd x \, x\, \sin(\pi
  x)\, \Gamma(\tfrac d2-x)\, \Gamma(\tfrac d2+x)\,.
  \label{generalized_free_energy}
\end{equation}
It was shown in \cite{Skvortsov:2017ldz} that the free energy of the
minimal type-A theory in AdS$_{d+1}$ was simply related to the above
quantity. On top of that, this expression is analytic in the $d$ and
therefore admits an extension to non-integer dimensions.  Below, we
will present another derivation of \eqref{generalized_free_energy}
from AdS and extend it to the case of the partially-massless
type-A$_\ell$ theories\,\footnote{Notice that the same result was
    obtained differently in \cite{Dowker:2010qy} for arbitrary
    dimensions and $\ell$.}.

Our derivation is based on the observation that the one-loop free
energy of the minimal type-A$_\ell$ theories coincides with that of
the $\rac_\ell$ singleton in AdS$_{d+1}$,
\begin{equation}
  \zeta_{\rm A_\ell^{min}}'(0) = \zeta_{\rac_\ell}'(0)\,,
\end{equation}
in all dimensions, as shown previously in \eqref{zetaracelladsodd} and
\eqref{rel_Amin_rac_odd}.  This field corresponds to the $so(2,d)$
module defined as the following quotient
\begin{equation}
  \mathcal D\big( \tfrac{d-2\ell}2; 0 \big) \cong \frac{ \mathcal V
    \big( \tfrac{d-2\ell}2; 0 \big)}{\mathcal V \big(
    \tfrac{d+2\ell}2; 0 \big)}\, ,
\end{equation}
and hence its zeta function in anti-de Sitter spacetime reads
\begin{equation}
  \zeta_{\rac_\ell}(z) = \zeta_{[\frac{d-2\ell}2;0]}(z) -
  \zeta_{[\frac{d+2\ell}2;0]}(z)\, .
\end{equation}
As recalled in \cite{Basile:2018zoy}, we can express the first
derivative of the AdS$_{d+1}$ zeta function as a spectral integral.
More precisely, for $d=2r$,
\begin{equation}
  \zeta_{[\Delta;\Y]}'(0) = -\ln R\, \int_0^{\Delta-\frac d2}\, \dd
  x\, \dim_{(-x-\frac d2,\Y)}^{so(d+2)} =
  -\zeta_{[d-\Delta;\Y]}'(0)\,,
\end{equation}
whereas for $d=2r+1$
\begin{equation}
  \zeta_{[\Delta;\Y]}'(0) - \zeta_{[d-\Delta;\Y]}'(0) = \pi\,
  \int_0^{\Delta-\frac d2}\, \dd x\, \tan(\pi\, x)\, \dim_{(-x-\frac
    d2,\Y)}^{so(d+2)}\,,
\end{equation}
for a bosonic representation. Applying the above expressions to the
$\rac_\ell$ singleton yields
\begin{itemize}
\item For $d=2r$,
  \begin{equation}
    \zeta_{[\frac{d-2\ell}2;0]}'(0) = \ln R \int_0^\ell \dd x\,
    \dim_{(-x-\frac d2,0)}^{so(d+2)} =
    -\zeta_{[\frac{d+2\ell}2,0]}'(0)\,,
  \end{equation}
  which leads to
  \begin{equation}
    \zeta_{\rac_\ell}'(0) = 2\, \ln R \int_0^\ell \dd x\,
    \dim_{(-x-\frac d2,0)}^{so(d+2)}\,;
  \end{equation}
\item For $d=2r+1$,
  \begin{equation}
    \zeta_{\rac_\ell}'(0) = -\pi\, \int_0^\ell \dd x\, \tan(\pi\, x)\,
    \dim_{(-x-\frac d2,0)}^{so(d+2)}\,.
  \end{equation}
\end{itemize}
One can recast the Weyl dimension formula involved in the above
integrals as
\begin{equation}
  \dim_{(-x-\frac d2,0)}^{so(d+2)} = (-1)^{r+1}\,\frac{2\,x}{\pi\,
    \Gamma(d+1)}\, \Gamma(\tfrac d2-x)\, \Gamma(\tfrac d2+x) \left\{
  \begin{aligned}
    \sin(\pi\,x)&\qquad\,\, [d=2r] \\ \cos(\pi\,x) &\qquad\,\,
        [d=2r+1]
  \end{aligned}
  \right.\,,
\end{equation}
so that we obtain
\begin{equation}
  \zeta_{\rac_\ell}'(0) = \frac{2\,v_d}{\Gamma(d+1)}\, \int_0^\ell\,
  \dd x\, x\, \sin(\pi\, x)\, \Gamma(\tfrac d2-x)\, \Gamma(\tfrac
  d2+x)\,,
  \label{ana zeta}
\end{equation}
with
\begin{equation}
  v_d = \left\{
  \begin{aligned}
    (-1)^{\frac{d}2+1}\,\frac{2}\pi\, \ln R\,& \qquad\,\, [d=2r]
    \\ (-1)^{\frac{d-1}2}\quad & \qquad\,\, [d=2r+1]
  \end{aligned}
  \right.\,.
\end{equation}
Note that \eqref{ana zeta} reproduces the generalized free energy
\eqref{generalized_free_energy} up to the factor $2\,v_d$, which
distinguishes even and odd $d$.\,\footnote{The integral in \eqref{ana
    zeta} is finite but the integrand diverges due to the poles of the
  Gamma function at $x-\frac d2\in \mathbb N$ which arise for
  $\ell>\frac{d}2$ and $d\notin 2\mathbb N$.  These poles are in fact
  responsible for the imaginary part of the free energy.}  It is
possible to unify the two cases and even extend it to any {\it real}
values of $d$ by replacing $v_d$ as
\begin{equation}
  v_d \ \to\ \tilde v_d=\frac1{\sin(\tfrac{\pi\,d}2)}\,,
  \label{v to tv}
\end{equation}
as was done in \cite{Giombi:2014xxa, Skvortsov:2017ldz}.  In the limit
$d$ goes to an odd integer, the new factor $\tilde v_d$ reproduces
$v_d$ without any divergence.  However, $\tilde v_d$ diverges in the
even $d$ limit. If we identify the pole with the factor $-\ln R$ in
$v_d$, then the residue correctly reproduces the other factor
$(-1)^{\frac d2}\,\frac2\pi$ in $v_d$. As explained in
\cite{Skvortsov:2017ldz}, the replacement \eqref{v to tv} amounts to
taking an alternative regularization for the AdS volume. Hence, the
zeta function \eqref{ana zeta} with $\tilde v_d$ (and the
corresponding one-loop free energy) reproduces the usual results for
any integer $d$ and is generalized to non-integer values of $d$.

\section{Type-B$_{\ell}$ higher-spin gravities}
\label{sec:typeBl}
We now turn to the holographic duality involving the type-B$_{\ell}$
higher-spin gravity. We will follow the discussion of the previous
section.

\subsection{AdS$_{2r+1}$}
\paragraph{Non-Minimal Theory}
We begin with the \textit{non-minimal} case.
Inserting the functions \eqref{etaxiBl} into \eqref{even d trick}, we
find that
\begin{equation}
  \zeta'_{{\rm B}_{\ell}^{\text{non-min}}}(0) = {\ln R}\, \oint
  \frac{\dd \b}{2\pi\,i}\, \oint \frac{\dd w}{2\pi\,i}\,
  \frac{4\sinh^2\big(\tfrac{2\ell-1}{2}{\beta}\big)\sinh w\,
    (\cosh\beta-\cosh w)}{(\beta^2-w^2) \cosh^2 \tfrac w2\, (\cosh
    w-1)^{r+1}\, (\cosh\beta - 1)^{r+1}}\,.
\end{equation}
For the same reason as in the type-A$_\ell$ case, i.e. the fact that
the integrand of the above integral is an even function of $\beta$, we
have
\begin{equation}
  \zeta_{{\rm B}^{\rm non-min}_\ell}'(0) = 0 \,,
\end{equation}
and hence the one-loop free energy vanishes for the non-minimal
type-B$_\ell$ theory.  Notice that the $\beta$ integrand behaves as
$e^{-\beta(\frac d2-2\ell+1)}$ when $\beta\rightarrow\infty$ and
therefore converges for $\ell<\tfrac{d+2}4$. As in the case of the
type-A$_\ell$ theory, this source of divergence can be traced back to
the fact that the Camporesi-Higuchi zeta function is singular for
$\bar \Delta=0$. Indeed, the scalar fields in the spectrum of the
type-B$_\ell$ theory have a minimal energy $\Delta$ given by
\begin{equation}
  \Delta = d-t-1\, , \qquad \qquad 0 \leqslant |t| \leqslant
  2(\ell-1)\, ,
\end{equation}
whereas for fields with spin-$(s,1^m)$ and $s\geqslant1$ this minimal
energy reads
\begin{equation}
  \Delta = s+d-t-1\, , \qquad \qquad 1 \leqslant t \leqslant
  2\ell-1\,,
\end{equation}
therefore in order for the spectrum to be devoid of fields with $\bar
\Delta$, one has to require $\ell<\tfrac{d+2}4$. We will consider the
analytic continuation in $\ell$ of the zeta function.

\paragraph{Minimal theory}
\label{par:min_typeBl}
We now turn to the minimal type-B$_\ell$ theory for which the
character is given by \eqref{chi-Bl-min}. The contribution of
the first term in \eqref{chi-Bl-min} has already been shown to
vanish, which leaves us with the contribution of the second term
alone.  Using \eqref{etaxiBlmin}, the relevant contour integral to be
computed, meaning the contribution of $-\tfrac12\,
\chi^{so(2,d)}_{\di_\ell}(2\beta;2\vec\alpha)$, reads
\begin{equation}
  \begin{split}
    \zeta{'}_{{\rm B}^{\rm min.}_\ell}\left(0\right) = - \frac{\ln
      R}{2^r} \oint {\dd\beta\over 2\pi i} \oint {\dd w\over 2\pi i}
         {\sinh\big((2\ell-1) \beta \big)\, \tanh w\,
           (\cosh\beta+\cosh w) \over (\beta^2-w^2)\, (\cosh w -
           1)^{r+1}\, (\cosh\beta + 1)^{r+1}}\,.
  \end{split}
\end{equation}
Again we carry out the $\beta$ integral and find that
\begin{equation}
  \begin{split}
    \zeta{'}_{{\rm B}^{\rm min.}_\ell}\left(0\right) = - \frac{\ln
      R}{2^{r-1}} \oint {\dd w \over 2\pi i\,w} {\sinh\big((2\ell-1) w
      \big) \over (\sinh w)^{2r+1}}\,,
  \end{split}
  \label{zeta_minBl_even}
\end{equation}
which reduces to a polynomial in $\ell$ of order $2r+1$ after
evaluation, as in the type-A$_\ell$ case.

\paragraph{Chiral type-B$_{\ell,\pm}$}
\label{par:chiral_typeBl}
For $d=2r$, one can consider a chiral $\di_\ell$ singleton, i.e.  Weyl
spinor carrying spin $(\tfrac12,\dots,\tfrac12,\pm\tfrac12)$ instead
of the direct sum of the two, namely Dirac spinor. The character of
such a conformal field reads
\begin{equation}
  \chi^{so(2,d)}_{\di_{\ell,\pm}}(\beta, \vec \alpha) =
  \sinh(\tfrac{2\ell-1}2\beta) \prod_{j=1}^r
  \frac{\cos\tfrac{\alpha_j}2}{\cosh\beta-\cos\alpha_j} \pm
  \cosh(\tfrac{2\ell-1}2\beta) \prod_{j=1}^r
  \frac{i\,\sin\tfrac{\alpha_j}2}{\cosh\beta-\cos\alpha_j}\,.
  \label{char_di_pm}
\end{equation}
The type-B$_{\ell,\pm}$ model or its minimal version is the higher-spin
theories whose spectrum are respectively given by the tensor product
or plethysm of the above character (see
\hyperref[app:chiralFF]{Appendix \ref{app:chiralFF}}). One can already
see that the computation of their zeta functions will only involve the
first part of \eqref{char_di_pm}. This is because  the second
term takes the form
\begin{equation}
  \eta(\beta) = \cosh(\tfrac{2\ell-1}2\beta)\,, \quad
  \xi(\beta,\alpha) = {i\sin\tfrac\alpha2 \over \cosh\beta -
    \cos\alpha}\,.
\end{equation}
Inserting this into \eqref{even d trick} we see that the resulting
contribution to the zeta function vanishes due to the fact that
$\xi\left(\beta,0\right)=0$.  Consequently, only the first term in
\eqref{char_di_pm} will contribute. This term is actually half of the
character of the parity-invariant $\di_\ell$ singleton used in the
previous computations, and hence we can conclude that
\begin{equation}
  \zeta'_{{\rm B}_{\ell,\pm}}(0) = 0\, , \qquad \quad \zeta_{{\rm
      B}_{\ell,\pm}^{\rm min.}}'(0) = \frac12\, \zeta_{{\rm
      B}_{\ell}^{\rm min.}}'(0)\, .
\end{equation}
This generalizes the result of \cite{Giombi:2016pvg}.

\paragraph{Order-$\ell$ Di module}
Paralleling the discussion in the previous section, let us compute the
one-loop free energy of the $\di_\ell$ singleton in
AdS$_{2r+1}$. Substituting \eqref{eq:char_di_ell} into \eqref{even d
  trick} yields
\begin{equation}
  \zeta_{\di_\ell}'(0) = 4\, \ln R\, \oint \frac{\dd \beta}{2\pi\,i}
  \oint \frac{\dd w}{2\pi\,i} \, \frac{\sinh
    \tfrac w2\,\sinh(\tfrac{2\ell-1}2\b)}{\big(\beta^2 - w^2\big)\, (\cosh
    w-1)^{r+1}}\,.
\end{equation}
After evaluating the $\beta$ integral, we end up with
\begin{equation}
  \zeta_{\di_\ell}'(0) = \frac{\ln R}{2^{r-1}}\, \oint \frac{\dd
    w}{2\pi\,i} \, \frac{\sinh(\tfrac{2\ell-1}2w)}{w\,
    (\sinh\tfrac w2)^{2r+1}}
\end{equation}
which is related to $\zeta_{\rm B_\ell^{min}}'(0)$ by a simple minus
sign (up to a rescaling of the integration variable of the above
contour integral).

\subsection{AdS$_{2r+2}$}
\paragraph{Non-Minimal Theory}
We next turn to the case of (non-minimal) type-B$_\ell$ theories in
even dimensional AdS space,
whose character is given by the square of the order-$\ell$
spin-$\tfrac12$ singleton $\chi^{so(2,d)}_{\di_\ell}(\beta;\vec
\alpha)$ defined in \eqref{eq:char_di_ell}. Using \eqref{etaxiBl} in
\eqref{odd d trick}, we find that the zeta function is given by
\begin{equation}
  \zeta_{{\rm B}_{\ell}^{\text{non-min}}}(z) = \frac
       {(-1)^r}{2^{2r+1}} \int_0^\infty \dd \beta\,
       \frac{\beta^{2z-1}}{\Gamma(2z)}\, \frac{\cosh\tfrac \b 2
         \,\sinh^2(\tfrac {2\ell-1}2\,\b)}{(\sinh\tfrac \b 2)^{2r+3}}\,.
\end{equation}
In terms of derivatives of the Lerch transcendent, the above zeta
function reads
\begin{equation}
  \begin{split}
    \zeta_{{\rm B}_{\ell}^{\text{non-min}}}(z) &=
    \frac{(-1)^{r}}{2\, (d+1)!}\, \partial_p^{d+1} \Big[
      \Phi(p,2z,-2\ell-r) + \Phi(p,2z,1-2\ell-r) \\& - 2
      \Phi(p,2z,-1-r)- 2\Phi(p,2z,-r)+\Phi(p,2z,-2+2\ell-r)\\ &\qquad+
      \Phi(p,2z,-1+2\ell-r)\Big]\Big\vert_{p=1}\,.
  \end{split}
  \label{B Lerc}
\end{equation}
The derivative of the above zeta function does not vanish, so it does
not follow the pattern of the holographic dualities of the other
higher-spin theories.  Moreover, by comparing the above
expression\footnote{Let us mention one subtlety in evaluating
  $\zeta_{{\rm B}_{\ell}^{\text{non-min}}}'(0)$ from \eqref{B
    Lerc}. The right hand side of the equality in \eqref{B Lerc} can
  be further expanded as a linear combination of $\zeta(2z-n,a)$ for
  some $n$ and $a$. However, the Hurwitz zeta function $\zeta(z,a)$ is
  not defined for Re$(z)>0$ and $-a\in\mathbb N$, and hence the
  derivative of the zeta function should be evaluated by taking the
  limit $z\rightarrow 0$ from the negative Re($z$).} with the CFT
results below, \eqref{zeta di ell d} and \eqref{eq:di_ell_even_lerch},
we do not find any simple relation between the one-loop free energies
of AdS and CFT.

\paragraph{Minimal Theory}
The zeta function of the minimal type-B$_\ell$ theory can be obtained
by adding the contribution of the second term in \eqref{chi-Bl-min} to
half of the non-minimal theory zeta function. This second contribution
can be computed by inserting \eqref{etaxiBlmin} into \eqref{odd d
  trick} so as to give
\begin{equation}
  \zeta_{\rm B_{\ell,2nd}^{min}}(z) = -\frac{1}{2^{r+2}}\,
  \int_0^\infty \frac{\dd\beta}{\Gamma(2z)}\, \beta^{2z-1}\,
  \frac{\sinh\big((2\ell-1)\beta\big)}{(\cosh\beta+1)^{r+1}}\,
  \oint\frac{\dd w}{2\pi\,i}\, \frac{\cosh\beta+w}{(\cosh\beta-w)\,
    w}\, \frac{1}{(w-1)^{r+1}}\,.
\end{equation}
The previous contour integral is given by the residue in $w=1$ of the
above integrand, which reads
\begin{equation}
  \oint\frac{\dd w}{2\pi\,i}\, \frac{\cosh\beta+w}{(\cosh\beta-w)\,
    w}\, \frac{1}{(w-1)^{r+1}} = (-1)^r +
  \frac{2}{(\cosh\beta-1)^{r+1}}\,,
  \label{B min w int}
\end{equation}
and hence we end up with
\begin{equation}
  \zeta_{\rm B_{\ell,2nd}^{min}}(z) = -\frac{1}{2^{2z+r+1}}\,
  \int_0^\infty \frac{\dd\beta}{\Gamma(2z)}\, \beta^{2z-1}\,
  \frac{\sinh\big(\tfrac{2\ell-1}2
    \beta\big)}{(\sinh\tfrac\beta2)^{2r+2}}\, \Big( 1+(-2)^r\,
  (\sinh\tfrac\beta4)^{2r+2} \Big)\,.
\end{equation}
As in the previous cases, one can recast the above expression into a
linear combination of derivatives of the Lerch transcendent, namely
\begin{eqnarray}
  \zeta_{\rm B_{\ell,2nd}^{min}}(z) & = & -\frac{2^{-2z+r}}{(2r+1)!}\,
  \Big(\frac{\partial}{\partial p}\Big)^{2r+1} \bigg[
    \Phi(p,2z,-r+\tfrac12-\ell) - \Phi(p,2z,-r-\tfrac12+\ell) \\ &&
    +\frac{(-1)^r}{2^{r+2}}\, \sum_{k=0}^{2r+2}\, (-1)^k
    \binom{2r+2}{k} \big( \Phi(p,2z,\tfrac{k-3r}2-\ell) -
    \Phi(p,2z,\tfrac{k-3r}2+\ell-1) \big)\bigg]_{p=1}\,.\nonumber
    \label{B min 2n}
\end{eqnarray}
This formula reproduces the previously obtained results
\cite{Gunaydin:2016amv, Giombi:2016pvg}, but unfortunately does not
seems to coincide with any CFT quantity.

\paragraph{Order-$\ell$ Di module}
To conclude the story in the AdS side, let us compute the zeta
function with the character of Di$_\ell$.  Substituting \eqref{di eta
  xi} into \eqref{odd d trick}, we obtain
\begin{equation}
  \zeta_{{\rm Di}_{\ell}}(z) =\int_0^\infty \frac{\dd
    \beta}{\Gamma(2z)}\, \oint \frac{\dd w}{2\pi\,i}
  \frac{\b^{2z-1}\,\sinh(\frac{2\ell-1}2\,\b)}{(\cosh\beta-
    w)\,(w-1)^{r+1}}\,.
\end{equation}
After the $w$ integral, it becomes
\begin{equation}
  \zeta_{{\rm Di}_{\ell}}(z) = \frac1{2^{r+1}}\int_0^\infty \frac{\dd
    \beta}{\Gamma(2z)}\, \frac{\b^{2z-1}\,
    \sinh(\frac{2\ell-1}2\,\b)}{(\sinh\frac\b2)^{2r+2}}\,.
\end{equation}
Notice that the above zeta function does not enjoy a relation like
\eqref{rel_Amin_rac_odd} of the Rac$_\ell$ case because of the second
term proportional to $(-2)^r$ in \eqref{B min 2n}.  Strangely, the
latter contribution can be removed by including the pole at $w=0$ in
\eqref{B min w int}.

\subsection{CFT$_d$}
\label{subsec:CFTfermion}
In the previous subsection, we have shown that the zeta function of
the non-minimal type-B$_\ell$ higher-spin gravity in AdS$_{2r+1}$ is
of order $\cO(z^2)$, which implies that its one-loop free energy
vanishes and therefore confirms the AdS/CFT duality reviewed
previously.  We were also able to obtain an integral expression for
the zeta function of the minimal theory, which coincides with that of
the order-$\ell$ Di module. We will relate this expression to that of
the $a$-anomaly coefficient of the $\di_\ell$ singleton in the
following subsection. Besides the even $d$ analysis, we will also
compute the free energy of the order-$\ell$ spin-$\tfrac12$ singleton
on the odd-dimensional sphere and thereby show that it is not simply
related to the (non-vanishing) one-loop free energy of the non-minimal
type-B$_\ell$ theory in AdS$_{2r+2}$ computed previously.

The order-$\ell$ spin-$\tfrac12$ singleton in $d$-dimensions is a free
conformal spinor field of conformal weight $\frac{d+1-2\ell}2$,
described by the action
\begin{equation}
  S_{\di_\ell}[\psi] = i\, \int \dd^d x\, \bar \psi\,
  \slashed{\partial}^{2\ell-1} \psi\, .
\end{equation} 
For Einstein manifolds, the extension of this order-$(2\ell-1)$ Dirac
operator was worked out in \cite{Dowker:2013mba, Fischmann2015} and in
the case of the $d$-dimensional sphere it can be factorized as
follows:
\begin{equation}
  \slashed \partial^{2\ell-1} \qquad \rightarrow \qquad
  \prod_{k=0}^{2(\ell-1)} \big( \slashed \nabla_{S^d} - (\ell-1-k)
  \big)\,,
\end{equation}
where $\slashed{\nabla}_{S^d}$ is the Dirac operator on the
$d$-sphere. The eigenvalues of $\slashed \nabla_{S^d}$ acting on a
Dirac spinor are $\pm(n+\tfrac d2)$ for $n\in\N$, where the sign $\pm$
refers to the upper and lower components of the spinor field
\cite{Camporesi:1995fb}. The eigenvalues to be considered in the
definition of the zeta function are therefore
\begin{equation}
  \lambda^{(\ell)}_{n,\pm} = \pm\, \prod_{k=0}^{2(\ell-1)}
  \lambda_{n,k}^{(\ell)}\, , \qquad \lambda_{n,k}^{(\ell)} :=
  \tfrac{d+1-2\ell}2+\tfrac12+n+k\, ,
\end{equation}
whose degeneracies are given by
\begin{equation}
  d_{\lambda^{(\ell)}_{n,\pm}} = \frac{2^r\, (n+d-1)!}{n!\, (d-1)!}
  \equiv \dim_{(n+\frac12,\frac12)}^{so(d+1)}\, .
\end{equation}
Notice that $(n+\tfrac12,\tfrac12)$ in the above equation denotes the
$so(d+1)$ irreducible representation defined by the highest weight
\begin{equation}
  (n+\tfrac12,\underbrace{\tfrac12,\dots,\tfrac12}_{r-1})\, , \qquad
  \text{for} \qquad d=2r,
\end{equation}
and
\begin{equation}
  (n+\tfrac12,\underbrace{\tfrac12,\dots,\tfrac12}_{r-1},\pm\tfrac12)\,
  , \qquad \text{for} \qquad d=2r+1\,.
\end{equation}
This leads to the following zeta function
\begin{equation}
  \zeta_{\di_\ell}^{(d)}(z) = 2\, \sum_{k=0}^{2(\ell-1)} \sum_{n=0}^\infty
  \dim_{(n+\frac12, \frac12)}^{so(d+1)}
      \left(\lambda^{(\ell)}_{n,k}\right)^{-z}\, .
\end{equation}
Notice that the overall factor $2$ in the above equation comes from
the fact that we take into account the contribution of both the 
upper and lower components, i.e. we hereafter consider a complex
spinor. Results for a Majorana spinor (available in $d=2,3,4,8,9$ mod
8) follow simply by dividing the quantities computed in this section
by $2$. The free energy of the $\di_\ell$ singleton is given by
\begin{equation}
  F_{\di_\ell} = -2\, \sum_{k=0}^{2(\ell-1)} \sum_{n=0}^\infty
  \dim_{(n + \frac12, \frac12)}^{so(d+1)}\, \ln
  \lambda^{(\ell)}_{n,k}\, ,
\end{equation}
and is therefore related to the zeta function \eqref{zeta_di} through:
\begin{equation}
  F_{\di_\ell} = {\zeta^{(d)}_{\di_\ell}}'(0)\, .
\end{equation}
Upon using
\begin{equation}
  \sum_{n=0}^\infty\, e^{-\beta n}\,
  \chi^{so(d+1)}_{(n+\frac12,\frac12)}(\vec \alpha)\big|_{\vec
    \alpha=\vec 0} = \Big[ \Pd{d+1} (i\beta; \vec \alpha)\,
    \chi_{\boldsymbol{\frac12}}^{so(d)}(\vec \alpha) \Big]_{\vec
    \alpha=\vec 0}\,,
\end{equation}
we can also express the zeta function \eqref{zeta_di} as a Mellin
transform:
\begin{eqnarray}
  \zeta_{\di_\ell}^{(d)}(z) & = & 2\, \sum_{k=0}^{2(\ell-1)}
  \sum_{n=0}^\infty \big[ \chi_{(n+\frac12,\frac12)}^{so(d+1)}(\vec
    \alpha) \big]_{\vec \alpha=\vec 0} \int_0^\infty \frac{\dd
    \beta}{\Gamma(z)}\, \beta^{z-1}\, e^{-\beta
    \lambda^{(\ell)}_{n,k}} \nonumber \\ & = & 2\, \int_0^\infty
  \frac{\dd \beta}{\Gamma(z)}\, \beta^{z-1}\,
  \frac{e^{-\beta/2}}{1-e^{-\beta}} e^{-\beta (\frac{d+1-2\ell}2)}
  (1-e^{-\beta(2\ell-1)}) \big[
    \chi^{so(d)}_{\boldsymbol{\frac12}}(\vec \alpha)\, \Pd d
    (i\beta;\vec \alpha) \big]_{\vec \alpha=\vec 0} \nonumber \\ & = &
  \int_0^\infty \frac{\dd \beta}{\Gamma(z)}\, \beta^{z-1}\,
  \frac{1}{\sinh\tfrac \beta 2}\, \chi_{{\rm
      Di}_\ell}^{so(2,d)}(\beta,\vec 0)\,.
  \label{zeta_di}
\end{eqnarray}
More explicitly, we have
\begin{equation}
  \zeta_{\di_\ell}^{(d)}(z) = \frac{1}{2^{d-r-1}}\, \int_0^\infty
  \frac{\dd \beta}{\Gamma(z)}\, \beta^{z-1}\,
  \frac{\sinh(\tfrac{2\ell-1}2\beta)}{(\sinh\tfrac\beta2)^{d+1}}\,.
  \label{zeta di ell d}
\end{equation}
Now using the Lerch transcendent, we can rewrite the above
  expression as 
\begin{equation}
  \zeta_{\di_\ell}^{(d)}(z) = \frac{2^{r+1}}{d!}\,
  \Big(\frac{\partial}{\partial p}\Big)^d\, \Big[
    \Phi(p,z,-\tfrac d 2+1-\ell) - \Phi(p,z,-\tfrac d2+\ell) \Big] \Big|_{p=1}
  \label{eq:di_ell_even_lerch}
\end{equation}
As in the $\rac_\ell$ singleton case, the above integral can be
divergent for two possible reasons. Firstly in the limit
$\beta\rightarrow\infty$, the integrand behaves as
$e^{-\beta(d+1-2\ell)/2}$ and as a consequence the integral is not
convergent when the conformal weight of the $\di_\ell$ singleton
becomes negative, i.e. when $\ell>\tfrac{d+1}2$. As in the scalar case
we resolve this issue by simply analytically continuing the zeta
function in $\ell$. Secondly, the integral \eqref{zeta_di} possesses a
pole at $\beta=0$, and this singularity can be handled as in the
scalar case. In addition, the character of the $\di_\ell$ singleton
obeys
\begin{equation}
  \chi_{\di_\ell}^{so(2,d)}(-\beta; \vec \alpha) = (-1)^{d+1}\,
  \chi^{so(2,d)}_{\di_\ell}(\beta; \vec \alpha)\,,
\end{equation}
and as a result the integrand of \eqref{zeta_di} for $z=0$ is odd/even
in even/odd dimensions, which in turn implies that it has a
non-vanishing residue only in even dimensions. This residue is related
the the conformal anomaly coefficient.

\subsubsection{CFT$_{2r}$}
\label{subsubsec:CFTfermion_even}
In analogy to the scalar case, the $a$-coefficient of the Weyl anomaly
of the $\di_\ell$ singleton on the $d$-sphere of radius $R$
(previously computed for $\ell=1$ in e.g. \cite{Cappelli:2000fe,
  Aros:2011iz}) corresponds to the coefficient of the $\ln R$ term in
the free energy, and hence it is related to the zeta function
\eqref{zeta_di} through
\begin{equation}
  a_{\di_\ell} = \zeta^{(2r)}_{\di_\ell}(0)\,.
\end{equation}
This coefficient is therefore given by the contour integral
\begin{equation}
  a_{\di_\ell} = \oint \frac{\dd \beta}{2\pi\,i\,\beta}\,
  \frac{\sinh(\tfrac{2\ell-1}2\beta)}{2^{d/2-1}\,
    (\sinh\tfrac\beta2)^{d+1}}\,.
  \label{anomaly_di}
\end{equation}
By comparing it with $\zeta_{\rm B^{min}_\ell}'(0)$, we see that the
two quantities are related through
\begin{equation}
  \zeta_{\rm B^{min}_\ell}'(0) = -\ln R\, \zeta_{\di_\ell}^{(2r)}(0) =
  -\ln R\, a_{\di_\ell}\,.
\end{equation}
Notice that the above relation implies that the one-loop free energy
of the minimal type-B$_\ell$ theory in AdS$_{2r+1}$ is given by {\it
  half} of the $a$-anomaly coefficient of the $\di_\ell$ singleton on
$S^{2r}$. As mentioned previously, this is a consequence of the fact
that we computed here the anomaly coefficient for a complex spinor. In
other words, the one-loop free energy of the minimal type-B$_\ell$
theory is given by the $a$-anomaly coefficient of a {\it Majorana}
spinor. Finally, let us point out that the residue \eqref{anomaly_di}
can be computed using the formula:
\begin{equation}
  a_{\di_\ell} = \frac{1}{(2r)!}\, \Big( \frac{\dd}{\dd\beta}
  \Big)^{2r} \Big[ \frac{\beta^{2r}}{2^{r-1}}\,
    \frac{\sinh(\tfrac{2\ell-1}2\beta)}{(\sinh\tfrac\beta2)^{2r+1}}
    \Big]\Big|_{\beta=0}\,.
\end{equation}
Some examples in low dimensions can be found in \hyperref[table
  fer]{Table \ref{table fer}}.
\begin{table}[h!]
  \centering
  \begin{center}
    \begin{tabular}{c|c}
      $d$ & $a_{\di_\ell}$\\[5pt] \hline\hline
      $2$ & $\frac13 (1-6\ell^2+4\ell^3)$\\[5pt]
      \hline
      $4$ & $-\frac{1}{90}(11 - 60 \ell^2 + 20 \ell^3 + 30 \ell^4 - 12
      \ell^5)$\\[5pt]
      \hline
      $6$ & $\frac{1}{3780} ( 191-1008\ell^2 + 224\ell^3 + 630\ell^4-
      168\ell^5 - 84\ell^6 +24\ell^7 )$\\[5pt]
      \hline 
      $8$ & $-\frac{1}{113400} (2497- 12960\ell^2 +2160\ell^3+
      8820\ell^4 $ \\[5pt] & \qquad $ -1764\ell^5- 1680\ell^6+
      360\ell^7 + 90\ell^8 -20\ell^9 )$ \\[5pt]
      \hline
      $10$ & $\frac{1}{7484400}(73985 - 380160 \ell^2+ 50688 \ell^3+
      270600 \ell^4 - 43296 \ell^5 $ \\[5pt] & $ - 60060 \ell^6 +
      10296 \ell^7+ 4950 \ell^8- 880 \ell^9- 132 \ell^{10} +24
      \ell^{11})$
    \end{tabular}
  \end{center}
  \caption{Summary of $a$-anomaly coefficients for the order-$\ell$
    Dirac spinor in low dimensions.}
  \label{table fer}
\end{table}

\subsubsection{CFT$_{2r+1}$}
\label{subsubsec:CFTfermion_odd}
By comparing the zeta function of the $\di_\ell$ singleton on
$S^{2r+1}$ given in \eqref{zeta_di} with the zeta function of the
non-minimal type-B$_\ell$ theory \eqref{zeta_minBl_even}, it is clear
that the free energy on both sides are unrelated.  This discrepancy
was already noticed in \cite{Giombi:2016pvg, Gunaydin:2016amv} for the
case $\ell=1$, and is therefore not surprising. Let us nevertheless
elaborate on a property of the free energy of the $\di_\ell$ CFT.
Similarly to the case of the scalar singleton, the conformal weight of
the order-$\ell$ spin-$\tfrac12$ singleton can become negative for
sufficiently large values of $\ell$, namely when $\ell>\tfrac{d+1}2$.
As a consequence, the free energy of the $\di_\ell$ singleton,
\begin{equation}
  F_{\di_\ell} = -2 \sum_{k=0}^{2(\ell-1)} \sum_{n=0}^\infty
  \dim_{(n+\frac12,\frac12)}^{so(d+1)} \, \ln\big(
  \tfrac{d+1-2\ell}2+\tfrac12+n+k\big),
\end{equation}
develops an imaginary part. For example, one finds for $d=3$
\begin{equation}
  F_{\di_\ell} = \frac{1}{16} \Big( -\tfrac23
  (2\ell-3)(2\ell-1)(2\ell+1)\, \ln 2 + 3(2\ell-1)
  \tfrac{\zeta(3)}{\pi^2}\Big) - \frac{i\pi}{6}\, (\ell-2)(\ell-1)\ell(\ell+1)\, .
\end{equation}
This imaginary part can be computed as in the scalar singleton
case. By introducing $m= \ell-r-2\,$ we are led to the sum
\begin{equation}
  i\,{\rm Im}( F_{\di_\ell} ) = -2i\,\pi\, \sum_{k=0}^{m}
  \sum_{n=0}^{m-k} \dim_{(n+\frac12,\frac12)}^{so(d+1)} = (-1)^r\,
  2^{r+1}\, \frac{i\pi}{(d+1)!}\, (1-\ell)_{r+1}\, (\ell)_{r+1}\, .
\end{equation}

\section{Generalizations of higher-spin theories}
\label{sec:genhigher-spin}

The type-A$_\ell$ and -B$_\ell$ higher-spin gravities can be
generalized to a few simply related models.

\subsection{Type-AB$_\ell$}

The spectrum of the non-minimal and minimal type-AB$_\ell$ theory can
be obtained by considering the ``weighted partition function'' of the
direct sum of the Rac$_\ell$ and Di$_\ell$ modules:
\begin{equation}
  Z_{{\rm DiRac}_{\ell}}
  :=\chi^{so(2,d)}_{\rac_\ell}-\chi^{so(2,d)}_{\di_\ell}\,.
\end{equation}
Note that the minus sign is not related to the submodule structure,
but the plethysm of fermionic modules.\footnote{We refer the reader to
  \cite{Bae:2017spv} for the appearance of the weighted partition
  function in the plethysm of fermionic modules.}  Then, the weighted
partition function of the non-minimal and minimal type-AB$_\ell$
theory reads
\begin{eqnarray}
  Z_{\rm AB^{\text{non-min}}_{\ell}}(\b,\vec \a) \eq \left[Z_{{\rm
        DiRac}_{\ell}}(\b, \vec \a)\right]^2 \nn \eq
  \chi^{so(2,d)}_{\rm A^{\text{non-min}}_\ell}(\b, \vec \a)
  +\chi^{so(2,d)}_{\rm B^{\text{non-min}}_\ell}(\b,\vec \a)
  -2\,\chi^{so(2,d)}_{\rac_{\ell}}(\beta, \vec\alpha)
  \chi^{so(2,d)}_{\di_\ell}(\beta, \vec\alpha)\,, \nn Z_{\rm AB^{\rm
      min}_{\ell}}(\b, \vec \a) \eq \frac12 \left[Z_{{\rm
        DiRac}_{\ell}}(\b, \vec \a)\right]^2 + \frac12\,Z_{{\rm
      DiRac}_{\ell}}(2\b, 2\vec \a) \nn \eq \chi^{so(2,d)}_{\rm A^{\rm
      min}_\ell}(\b, \vec \a) + \chi^{so(2,d)}_{\rm B^{\rm
      min}_\ell}(\b, \vec \a) -\chi^{so(2,d)}_{\rac_\ell}(\beta,
  \vec\alpha) \chi^{so(2,d)}_{\di_\ell}(\beta, \vec\alpha).
\end{eqnarray}
Therefore, to compute the zeta function for the type-AB$_\ell$ theory
it is sufficient to add the contribution of the term
$\chi^{so(2,d)}_{\rac_{\ell}}\left(\beta,\vec{\alpha}\right)
\chi^{so(2,d)}_{\di_\ell}\left(\beta,\vec{\alpha}\right)$ to the zeta
function $\zeta_{{\rm A}^{\text{(non-)min}}_\ell}+\zeta_{{\rm
    B}^{\text{(non-)min}}_\ell}$.  This is the contribution
corresponding to the fermionic partially-massless fields of depth
$t=1,\dots,2\ell-1$, and hence one should compute the zeta function
with the fermionic measure in AdS$_{2r+2}$ (i.e. using \eqref{f 1 even
  AdS} with $\epsilon=-1$).
\begin{itemize}
\item In AdS$_{2r+1}$, the contribution of the fermionic tower of
  partially-massless higher-spin fields to $\zeta_{\rm AB_\ell}'(0)$
  is proportional to
  \begin{equation}
    \oint \frac{\dd\beta}{2i\,\pi}\, \frac{\sinh(\ell\beta)\,
      \sinh(\tfrac{2\ell-1}2\beta)}{(\cosh\beta - 1)^{r+1}}\,\oint
    \frac{\dd w}{2i\,\pi} \, \frac{ \sinh\tfrac w2\, (\cosh\beta-\cosh
      w)}{(\beta^2-w^2)\, (\cosh w-1)^{r+1}}\,.
  \end{equation}
  The integrand of the above integral being an even function of
  $\beta$, this contribution identically vanishes.
\item In AdS$_{2r+2}$ the contribution of the tower of fermionic
  fields to $\zeta_{\rm AB_\ell}'(0)$ is proportional to (using the
  fermionic measure for the zeta function \eqref{trick_odd})
  \begin{equation}
    \frac{1}{(\cosh\beta-1)^{r+1}}\, \oint \frac{\dd w}{2i\,\pi}
    \frac{1}{(w-1)^{r+1}} = 0\, ,
  \end{equation}
  and hence this contribution also identically vanishes.
\end{itemize}
Therefore, we can see that the tower of fermionic fields does not
contribute to the zeta function of the type-AB$_\ell$ theory in any
dimensions. The same fact was obtained for the $\ell=1$ case in
\cite{Giombi:2016pvg, Gunaydin:2016amv}.

\subsection{Higher power of Rac$_\ell$}
Here we consider the higher-spin theory whose spectrum is given by the
tensor product of $n$ order-$\ell$ scalar singletons (that we will
denote type-A$_\ell^n$). Its character therefore reads
\begin{equation}
  \chi^{so(2,d)}_{{\rm A}^n_\ell}(\beta, \vec\alpha) = \left[
    \chi^{so(2,d)}_{\rac_\ell}(\beta, \vec\alpha) \right]^n\,.
\end{equation}
This spectrum corresponds to that of the $n$-linear operators on the
boundary and may be considered to be multi-particle states in
higher-spin gravity or the states in higher Regge trajectory in a
string-like theory dual to a matrix model CFT.  For such a character,
we can use the trick introduced in \eqref{trick_even} and
\eqref{trick_odd} with
\begin{equation} 
  \eta_{{\rm A}^n_\ell}(\b)=\frac{\sinh^n(\ell\,\b)}{2^{n(d-1-r)}\,
    (\sinh\frac\beta2)^{n(d-2r)}} ,\qquad \xi_{{\rm
      A}^n_\ell}(\b,\a)=\frac1{(\cosh\b-\cos\a)^n}\,.
\end{equation}

\subsubsection*{AdS$_{2r+1}$}\label{racell n}
Using the expression \eqref{even d trick} with the above function
yields
\begin{equation}
  \zeta_{{\rm A}_\ell^n}'(0) = \frac{\ln R}{2^{n(r-1)}}\, \oint
  \frac{\dd\beta}{2\pi\, i}\, \oint \frac{\dd w}{2\pi\,i} \, \frac{
    \sinh^n(\ell\beta)\, \sinh w\, (\cosh\beta - \cosh
    w)^{n-1}}{(\beta^2-w^2)\, (\cosh\b - \cosh w)^{(n-1)(r+1)}\,
    (\cosh w-1)^{r+1}}\,.
  \label{f_Anl_even}
\end{equation}
Due to the fact that both \eqref{f_Anl_even} and $\eta_{{\rm
    A}^n_\ell}$ are even functions of $\beta$ when $n$ is even, their
product does not have any residue at $\beta=0$ and therefore one can
conclude that
\begin{equation}
  \zeta_{{\rm A}^n_\ell}'(0) = 0\, , \qquad \text{for} \qquad n \in
  2\,\N\,.
\end{equation}
In particular, the usual non-minimal type-A$_\ell$ theory, which
corresponds to the case $n=2$, falls into this category.

\subsubsection*{AdS$_{2r+2}$}
Using the trick \eqref{trick_odd} with $\xi_{{\rm
    A}^n_\ell}(\beta,\alpha)$ produces the following contour integral
\begin{equation}
  \frac{1}{(\cosh\beta-1)^{(r+1)(n-1)}} \oint \frac{\dd w}{2i\pi}
  \frac{(\cosh\beta-w)^{n-2}}{(w-1)^{r+1}} = \frac{(2-n)_r}{r!}
  \frac{1}{(\cosh\beta-1)^{nr+1}}\, ,
\end{equation}
so that the zeta function for the type-A$^n_\ell$ theory is given by
\begin{equation}
  \zeta_{{\rm A}_\ell^n}(z) = \frac{(2-n)_r}{2^{2(nr+1)}\,
    r!}  \int_0^\infty \frac{\dd\beta}{\Gamma(2z)} \beta^{2z-1}\,
  \frac{\sinh\beta\, \sinh^n(\ell\beta)}{(\sinh\tfrac\beta2)^{nd+2}}\, .
\end{equation}
The Pochhammer symbol in the above expression ensures that
\begin{equation}
  \zeta_{{\rm A}_\ell^n}(z) = 0 \qquad \text{for} \qquad 2
  \leqslant n \leqslant r+1\, .
\end{equation}
In particular, we recover that the partially-massless type-A$_\ell$
theories ($n=2$) have a vanishing free energy in all dimensions as we
observed in the previous section.

\subsection{A Stringy Version of Type A$_\ell$ dualities}
Let us now briefly turn our attention to the free $SU(N)$ matrix model
CFT with a Rac$_\ell$ scalar in $d=2r$, treated for $\ell=1$ and $d=4$
(as well as $d=3$) in \cite{Bae:2016rgm}. Our discussion follows that
paper quite closely. The reader may consult \cite{Bae:2017fcs} for a
review.

For this model, the spectrum the theory is given by the direct sum of
the cyclic tensor product of $n$ $\rac_\ell$ modules (denoted
$\mathrm{cyc}^n$) for $n\geqslant2$. The relevant character of the
$n$th cyclic tensor product of $\rac_\ell$ reads
\begin{equation}\label{cyc n}
  \chi^{so(2,d)}_{\mathrm{cyc}^n}(\beta, \vec\alpha)= \frac1n\,
  \sum_{k|n} \varphi(k) \left[ \chi^{so(2,d)}_{\rac_\ell}(k\beta,\,
    k\vec\alpha) \right]^{n/k}\,,
\end{equation}
where the notation $k|n$ indicates that $k$ is a divisor of $n$. The
$n=2$ case corresponds to the partition function of the minimal type-A$_\ell$ higher-spin theory already considered above. Let us now turn
to the higher $n$'s.

The contribution of $\mathrm{cyc}^n$ to the first derivative of the
zeta function is given by the contour integrals
\begin{eqnarray}
  \zeta_{\mathrm{cyc}^n}'(0) & = & \ln R\, \oint \frac{\dd
    \beta}{2i\,\pi}\, \oint \frac{\dd w}{2i\,\pi}\,
  \frac{\sinh^{n/k}(k\ell\beta)\, (\cosh k\beta - \cosh k
    w)^{n/k}}{(2^{r-1} [\cosh
      k\beta-1]^{r+1})^{n/k}} \label{integral_cyclic} \\ && \qquad
  \qquad \qquad\qquad \times \frac{\sinh
    w}{(\beta^2-w^2)(\cosh\beta-\cosh w)}\,
  \Big(\frac{\cosh\beta-1}{\cosh w-1}\Big)^{r+1}\,. \nonumber
\end{eqnarray}
As in the previously studied cases, the $\beta$ integral is easier to
perform first. The potential poles are at $\beta=\pm w$ and at $\beta=0$,
but their contribution does not vanish only when certain conditions are met.
\begin{itemize}
\item To examine the point $\beta=\pm w$, it is sufficient to consider the following part of  \eqref{integral_cyclic}: 
 \begin{equation}
 \frac1{\beta^2 - w^2}\,
    \frac{(\cosh k\beta - \cosh k w)^{n/k}}{\cosh\beta - \cosh w}\,.
  \end{equation}
Due to the first factor the above has a pole at $\beta=\pm w$  unless the second factor has a zero at the same point.
This happens when $n/k\ge 2$, that is, unless $k=n$.
To repeat, the $\beta$ integral of   \eqref{integral_cyclic}
receives the contribution from the pole at $\beta=\pm w$  if and only if $k=n$.

\item If $n/k$ is an even integer, 
then the integrand of \eqref{integral_cyclic} becomes an even function of $\beta$
which is free of pole at $\beta=0$.
Hence, the contribution from the pole at $\beta=0$ can arise only for odd $n/k$.
Yet when $n/k=1$, the pole disappears again due to the zero of the numerator at $\beta=0$.
\end{itemize}
While the contribution coming from the pole in $\beta=0$ is quite
difficult to extract in full generality, the contribution  from
the poles in $\beta=\pm w$ can be computed in arbitrary dimensions.
In this case, the contour integral to perform reads
\begin{equation}
    \frac{\ln R}{2^{r-1}}\, \frac{\varphi(n)}n\, \oint {d\beta\over
      2\pi i} \oint {dw\over 2\pi i} \,\frac{\sinh w\,\sinh
      n\ell\beta\, (\cosh n\beta -\cosh n w)}{(w^2-\beta ^2) (\cosh
      n\beta -1)^{r+1} (\cosh w -
      \cosh\beta)}\,.
\end{equation}
As in Equation \eqref{minzeta 1}, we carry out the $\beta$ integral
first by picking up the poles at $\beta = \pm w$. We find
\begin{equation}
  \begin{split}
   & \frac{\ln R}{2^{r-1}}\, \varphi(n)\,\oint {dw\over 2\pi i\,
      w}\, \frac{\sinh n w\, \sinh(\ell n w)}{2^{r+1}\, (\sinh n \frac
      w2)^{2r+2}} \\ & = \varphi(n) \, \frac{\ln R}{2^{2r-1}}\,
    \oint {dw\over 2\pi i\, w} \frac{\cosh \frac w2\, \sinh \ell
      w}{(\sinh\frac w2)^{2r+1}} = \varphi(n)\,
    \zeta^{(2r)}_{\rac_\ell}(0)\,,
  \end{split}
  \label{contribution_n=k}
\end{equation}
where we have rescaled $w$ in the last step to make contact with
\eqref{zeta d even}.

Notice that when $n=2^m$ for an integer $m$, the divisors $k$ of $n$
are $k=2^p$ for $0 \leqslant p \leqslant m$, so that the only odd integer $n/k$ is 1. 
According to the previous discussion, the sum over $k$
in \eqref{cyc n} then reduces to the term $k=n$, and the computation of
$\zeta_{\mathrm{cyc}^n}'(0)$ boils down to the contribution of
\eqref{contribution_n=k}. 
In this way, we prove that
\begin{equation}
  \Gamma^{\sst (1)}_{\mathrm{cyc}^n} = \varphi(n)\,
  \Gamma^{\sst (1)}_{\rac_\ell} \qquad [n=2^m]\,.
  \label{euler id}
\end{equation}
This behavior was first observed in \cite{Bae:2016rgm} for $m=1$ to 5 in $\ell=1$ and $d=4$. 
Our analysis provides a proof of this for arbitrary $\ell$, $d=2r$ and $m$.

As mentioned above, evaluation of the contour integral form of CIRZ is generically 
complicated because the $\beta=0$ pole can contribute. For this reason, to evaluate
the free energy contribution from $\mathrm{cyc}^n$ for generic $n$, we fix $d=4$ and 
use the derivative
form of CIRZ obtained in
\cite{Bae:2016rgm}, which reads (in the notations of
\cite{Basile:2018zoy})
\begin{equation}\label{dercirz1}
  \zeta'_{\cal H}(0) = \ln\,R\, \oint \frac{\dd\beta}{2\,\pi\,i}\,
  \sum_{n=0}^2\, (-1)^n\, \frac{2^{2n+1}\,n!}{\beta^{2(n+1)}}\,
  \mathsf{f}_{\cal H}^{4,(n)}(\b)\,,
\end{equation}
with
\begin{eqnarray}\label{dercirz2}
  \mathsf{f}_{\cal H}^{4,(0)}(\beta) & = & \Big[ 1 -
    \sinh^2\tfrac\beta2\, (\tfrac13\sinh^2 \tfrac \b
    2-1)\,(\partial_{1}^2 + \partial_{2}^2) \\ && \quad - \tfrac13\,
    \sinh^4\tfrac\beta2\, \big(\partial_{1}^4+ \partial_{2}^4 -12\,
    \partial_{1}^2\,\partial_{2}^2 \big) \Big]\, \chi^{so(2,4)}_{\cal
    H}(\beta,\vec\alpha)\big|_{\vec\alpha=\vec0}\,\,, \nonumber
\end{eqnarray}
\begin{equation}\label{dercirz3}
  \mathsf{f}_{\cal H}^{4,(1)}(\beta) = 
  \sinh^2\tfrac\beta2\, \Big[ \tfrac 1 3 \sinh^2 \tfrac \b 2-1 - 
    \sinh^2\tfrac\beta2\,(\partial_{1}^2 + \partial_{2}^2) \Big]\,
  \chi^{so(2,4)}_{\cal
    H}(\beta,\vec\alpha)\big|_{\vec\alpha=\vec0}\,\,,
\end{equation}
and
\begin{equation}\label{dercirz4}
  \mathsf{f}_{\cal H}^{4,(2)}(\beta) = \tfrac12\,
  \sinh^4\tfrac\beta2\, \chi^{so(2,4)}_{\cal H}(\beta, \vec0)\,.
\end{equation}
After straightforward computations for $n=3,\, 4$, we find
\begin{equation}\label{cyc 3}
  \Gamma^{\sst (1)}_{\mathrm{cyc}^3} = \frac{\ell ^3 \left(4026\, \ell
    ^8-20500\, \ell ^6 +37128\, \ell ^4-572118\, \ell
    ^2+914375\right)}{16329600}\,\ln R\,,
\end{equation}
and
\begin{equation}
  \Gamma^{\sst (1)}_{\mathrm{cyc}^4} = \frac{1}{90}\, \ell ^3 \left(5-3 \ell
  ^2\right)\, \ln R\,.
\end{equation}
In general, expressions for the vacuum energy
$\Gamma^{\sst (1)}_{\mathrm{cyc}^n}$ are rather complicated functions of
$\ell$, for instance $\Gamma^{\sst (1)}_{\mathrm{cyc}^5}$ is already a
polynomial of order 21 in $\ell$. This can be traced back to the fact
whenever the order $n$ has divisors $k$ such that $n/k$ is odd, the
pole in $\beta=0$ of \eqref{integral_cyclic} contributes, and the
order of this pole depends on $n$ and $k$. As a consequence, the order
of the resulting polynomial $\ell$ grows with these parameters. 
\begin{figure}[h]
  \centering
  \begin{minipage}{7cm}
    \includegraphics[width=6.5cm]{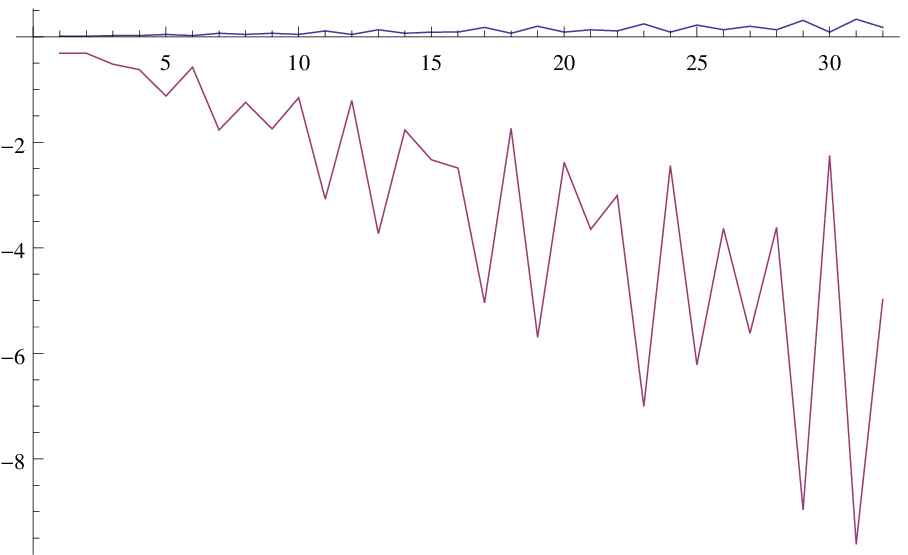}
    \caption{\footnotesize 
      $\Gamma^{\sst (1)}_{\mathrm{cyc}^n}$ plotted in units of $\log R$,
      from $n=1$ to 32. The blue line is for $\ell=1$ while the purple
      line is for $\ell=2$.}
    \label{fig:test1}
    \end{minipage}
  \qquad
  \begin{minipage}{7cm}
    \includegraphics[width=6.5cm]{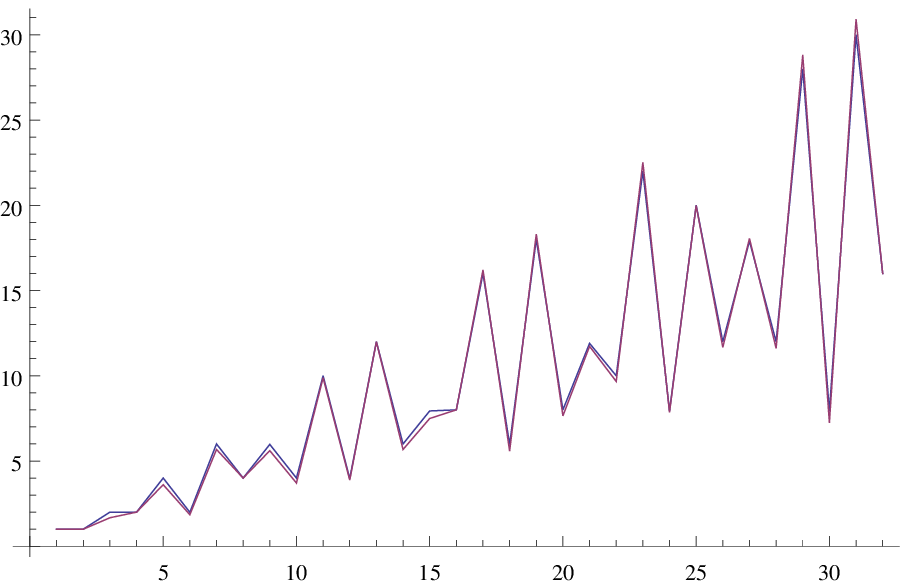}
     \caption{\footnotesize 
      $\Gamma^{\sst (1)}_{\mathrm{cyc}^n}$ plotted in units of
      $\Gamma^{\sst (1)}_{\mathrm{Rac}_\ell}$, from $n=1$ to 32. The blue
      line is for $\ell=1$ while the purple line is for $\ell=2$.}
    \label{fig:test2}
  \end{minipage}
\end{figure}
In
order to better illustrates the behavior of the one-loop free energy
of $\mathrm{cyc}^n$ with respect to $n$ and $\ell$, we also display in
\hyperref[fig:test1]{Figure \ref{fig:test1}} a plot of
$\Gamma^{\sst (1)}_{\mathrm{cyc}^n}$ for $\ell=1$ and $\ell=2$ for $n=1$ to
$32$ (plotted in units of $\ln R$). 
We see that the two curves are
quite well separated, with the magnitudes of
$\Gamma^{\sst (1)}_{\mathrm{cyc}^n}$ being much larger for $\ell=2$ than
those for $\ell=1$. At first sight this simply reflects the high
sensitivity to $\ell$ in these expressions, already visible in
\eqref{cyc 3}.
On the other hand, if we consider the behaviour of
${\Gamma^{\sst (1)}_{\mathrm{cyc}^n}/
  \Gamma^{\sst (1)}_{\mathrm{Rac}_\ell}}$ for $\ell=1$ and $\ell=2$,
displayed in \hyperref[fig:test2]{Figure \ref{fig:test2}}, we see that
the two graphs almost coincide with each other.

Finally, we turn to the vacuum energy contribution from the full
stringy spectrum, i.e. the spectrum encoded in the character obtained
from summing \eqref{cyc n} from $n=2$ to $\infty$. The resulting
character is given by \cite{Sundborg:1999ue}
\begin{equation}
  \chi^{so(2,4)}_{SU(N);\rac_\ell}(\beta,\,\vec{\alpha}) =
  -\chi^{so(2,4)}_{\rac_\ell}(\beta,\,\vec{\alpha}) -
  \sum_{k=1}^{\infty}{\varphi(k)\over k}
  \log\left[1-\chi^{so(2,4)}_{\rac_\ell}(k\beta,\,k\vec{\alpha})\right]
  \,.
\end{equation}
Applying the formulae \eqref{dercirz2}, \eqref{dercirz3} and \eqref{dercirz4} to
\begin{equation}
  \chi_{\cal H}^{so(2,4)}(\beta, \vec\alpha) =
  \chi_{\log,k}^{so(2,4)}(\beta, \vec\alpha) = -
  \log\left[1-\chi^{so(2,4)}_{\rac_\ell}(k\beta,\,k\vec{\alpha})\right]\,,
\end{equation}
yields
\begin{equation}
  \mathsf{f}_{\log,k}^{4,(0)}(\beta) = -\frac12\,\sinh^4
  \tfrac\beta2\, \log\big[ 1-
    \frac{\sinh(k\ell\beta)}{8\,\sinh^4\tfrac{k\beta}2} \big]\,,
\end{equation}
\begin{equation}
  \begin{split}
    \mathsf{f}_{\log,k}^{4,(1)}(\beta) = -\frac13\,
    \sinh^2\tfrac\beta2\, \big(\sinh^2\tfrac\beta2 - 3)\,& \log\big[
      1- \frac{\sinh(k\ell\beta)}{8\,\sinh^4\tfrac{k\beta}2} \big]\,
    \\ & + \frac{k^2 \, \sinh^4\tfrac\beta2\,
      \sinh(k\ell\beta)}{\sinh^2\tfrac{k\beta}2\, \big(8\sinh^4
      \tfrac{k\beta}2 - \sinh(k\ell\beta)\big)}\,,
  \end{split}
\end{equation}
\begin{equation}
  \begin{split}
    \mathsf{f}_{\log,k}^{4,(2)}(\beta) & = -\log\big[ 1-
      \frac{\sinh(k\ell\beta)}{8\,\sinh^4\tfrac{k\beta}2} \big]\, +
    \frac{k^4 \, \sinh^4\tfrac\beta2\, \sinh^2(k\ell\beta)}{2\,
      \sinh^4\tfrac{k\beta}2\, \big(8\sinh^4 \tfrac{k\beta}2 -
      \sinh(k\ell\beta)\big)^2}\, \\ & \qquad - \frac13 k^2 \big(
    (k^2-1) \sinh^2\tfrac\beta2 + 3\big)\frac{\sinh^2\tfrac\beta2\,
      \sinh(k\ell\beta)}{\sinh^2\tfrac{k\beta}2\, \big(8\sinh^4
      \tfrac{k\beta}2 - \sinh(k\ell\beta)\big)}\,,
  \end{split}
\end{equation}
One can check that by expanding the above expression around $\beta=0$
they are devoid of terms of order $\beta^1$, $\beta^3$ and
$\beta^5$ respectively. Hence, by virtue of \eqref{dercirz1}, 
$\chi_{\log,k}^{so(2,4)}$ never
contributes to the one-loop free energy of the $SU(N)$ matrix model. As
a consequence, we eventually find that
\begin{equation}
  \Gamma^{(1)}_{SU(N);\rac_\ell} = -\Gamma^{(1)}_{\rac_\ell}\,.
\end{equation}
Notice finally that some tacit assumptions in this prescription are
discussed in \cite{Bae:2016rgm,Bae:2017spv}.

\section{Discussion}
\label{sec:discussion}
In this paper we have applied the arbitrary dimensional CIRZ formula
obtained in \cite{Basile:2018zoy} to partially-massless higher-spin
gravities. Firstly, we found that all the theories considered in this
paper do not have any UV divergence in its one-loop free energy.
Concerning the finite part, the non-minimal type-A$_\ell$ theories in
$d+1$ dimensions have
\begin{equation}\label{gamma_nonmin_al}
  \Gamma^{\sst (1)}_{{\rm A}_{\ell}^{\text{non-min}}}=0\,,
\end{equation}
which is consistent with the CFT expectation.  On the other hand, the
minimal type-A$_\ell$ theories have
\begin{equation}\label{gamma_min_al}
  \Gamma^{\sst (1)}_{{\rm A}_{\ell}^{\text{min}}}=
  \left\{
  \begin{array}{cc} 
    \ln R\,a_{\rac_\ell} & \qquad [d=2r]\\ F_{\rac_\ell} & \qquad
        [d=2r+1]
  \end{array}\right..
\end{equation}
Concerning the type-B$_\ell$ theories, we find an analogous result for
even $d$
\begin{equation} \label{gamma_min_bl_odd}
  \Gamma^{\sst (1)}_{{\rm B}_{\ell}^{\text{non-min}}}=0\,, \qquad
  \Gamma^{\sst (1)}_{{\rm B}_{\ell}^{\text{min}}} = \tfrac12\, \ln R\,
  a_{\di_\ell}\,,
\end{equation}
but for odd $d$, we do not find a relation between the AdS one-loop
free energy and the free energy of the order-$\ell$ free fermion.  These results have been
obtained previously in the $\ell=1$ case \cite{Giombi:2013fka,
  Giombi:2014iua} and for the type-A$_2$ case for $d$ up to 20
\cite{Brust:2016xif}.  The results about the minimal theories may fit
in with the holographic conjecture by introducing a shift in the
dictionary between the bulk coupling constant $g$ and the number of
conformal fields $N$ \cite{Giombi:2013fka, Giombi:2014iua}: $g^{-1} =
N-1\,.$ Meanwhile, the results for the putative stringy dualities
seem to suggest the relation $g^{-1} =
N^2$ \cite{Bae:2016rgm}.
We suggest the same interpretation for our results, which have
been derived for arbitrary $d$ and $\ell$.

In obtaining our results for partially-massless higher-spin theories using CIRZ, we could reconfirm that the zeta function regularization
renders finite not only the divergences from UV but also those from the sum over spectrum.
However, as we are considering non-unitary theories,
several signs of non-unitarity show up in the form of IR divergences.
They arise in the $\b$ integral for the large $\b$ region.
The parameter $\b$ has a clear meaning of the inverse energy scale, as one can see
from its role in the character: small $\b$ corresponds to  high energy
and large $\b$ to small energy. 
The lowest energy of the theory decreases as $\ell$ increases,
and we could see that the $\b$ integral starts to diverge for $\ell\ge \frac{d}4$ and $\ell\ge \frac{d+2}4$
in the type-A$_\ell$ and type-B$_\ell$ theories, respectively.
 From the AdS perspective, the divergences arising for higher $\ell$ 
 are caused by the fields with vanishing $\bar\D=\D-\frac d2$,
 which can be interpreted as the AdS counterpart of the IR divergence of the massless fields in flat spacetime.
 This kind of IR divergence could be removed by an analytic continuation in $\ell$.
 As $\ell$ increases further, the theory contains fields with negative 
 $\D$ (hence negative energy states)
 for  $\ell\ge \frac{d}2$ and $\ell\ge \frac{d+1}2$ in type-A$_\ell$ and type-B$_\ell$ theories, respectively.
 Interestingly, these bounds correspond to that for the IR divergence of the CFT,
 meaning the divergence of the CFT zeta function in the large $\b$ region. 
Above this bound, the even $d$ free energy develops an imaginary part,
which could be exactly calculated.
It would be interesting to better understand the physical implications of these issues
for both AdS and CFT sides.

\acknowledgments

It is a pleasure to thank Evgeny Skvortsov 
and Stuart Dowker
for the interesting and useful correspondence. We are especially grateful to S. Dowker for pointing out a typo in an earlier version of this paper.
  The research of T.B., E.J. and W.L.  was supported by the National
  Research Foundation (Korea) through the grant 2014R1A6A3A04056670.
  S.L.'s work is supported by the Simons Foundation grant 488637 
  (Simons Collaboration on the Non-perturbative bootstrap)
  and the project CERN/FIS-PAR/0019/2017.
  Centro de Fisica do Porto is partially funded by the 
  Foundation for Science and Technology of Portugal (FCT).

\appendix

\section{Type-B$_\ell$ minimal model}
\label{app:minBl}
In this appendix, we spell out the decomposition of the
antisymmetrized tensor product of two $\di_\ell$ singletons in
arbitrary dimensions, which corresponds to the spectrum of the minimal
type-B$_\ell$ theory. To do so, we will use the previously introduced
characters expressed in terms of the variables
\begin{equation}
  q := e^{-\beta}\,, \qquad \text{and} \qquad x_k := e^{i\alpha_k}\,,
  \quad k=1,\dots,r\,,
\end{equation}
instead of $\beta$ and $\vec\alpha$. For instance, the character of
the $\di_\ell$ singleton then reads
\begin{equation}
  \chi^{so(2,d)}_{\di_\ell}(q, \bar x) = q^{\frac{d+1-2\ell}2}\,
  (1-q^{2\ell-1})\, \chi^{so(d)}_{\boldsymbol{\frac12}}(\bar x)\, \Pd
  d (q, \bar x)\,,
\end{equation}
with $\bar x = (x_1, \dots, x_r)$ and
\begin{equation}
  \Pd d (q, \bar x) = \frac{1}{(1-q)^{d-2r}}\, \prod_{k=1}^r\,
  \frac{1}{(1-qx_k)(1-qx_k^{-1})}\,.
\end{equation}
Characters of the $so(d)$ algebra can be found in
\cite{Basile:2018zoy} or the textbooks \cite{Weyl1939,
  Fulton1991}. For more details on character of the conformal algebra,
see e.g. \cite{Dolan:2005wy, Beccaria:2014jxaw, Bourget:2017kik}.

\subsection{Odd $d$}
\label{app:minBlodd}
In order to decompose the plethysm of two $\di_\ell$ singletons using characters, 
we need to decompose
\begin{equation}
  \frac12\, \big( \chi^{so(2,d)}_{\di_\ell}(q, \bar x) \big)^2 \pm
  \frac12\, \chi^{so(2,d)}_{\di_\ell}(q^2, \bar x^2)
\end{equation}
into a sum of characters of irreducible $so(2,d)$ modules. Knowing the
decomposition of the tensor product of two $\di_\ell$ singletons,
i.e. the first term in the above equation, we can simply focus on the
second term. To deal with it, we will need a few key identities, namely:
\begin{itemize}
\item The $\Pd d$ function evaluated in $(q^2, x_1^2, \dots, x_r^2)$ can be
  factorized into a product of two such functions
  \begin{equation}
    \Pd d (q^2, \bar x^2) = \Pd d (q, \bar x)\, \Pd d (-q, \bar x)\,,
  \end{equation}
  which can in turn be expanded as the series
  \begin{equation}
    \Pd d (\pm q, \bar x) = \frac{1}{1-q^2}\, \sum_{s=0}^\infty (\pm
    q)^s\, \chi_{(s)}^{so(d)}(\bar x)\,.
  \end{equation}
\item The character of the spin-$\tfrac12$ $so(d)$ representation can
  be expanded as
  \begin{equation}
    \chi^{so(d)}_{\boldsymbol{\frac12}}(\bar x^2) = \sum_{m=0}^r \epsilon_m
    \chi^{so(d)}_{(1^{r-m})}(\bar x)\, , \qquad \epsilon_m :=
    (-1)^{m(m+1)/2}\,.
  \end{equation}
\item Finally, the product of two $so(d)$ characters can be decomposed according to the 
  tensor product rule.
\end{itemize}
Using the above properties, one ends up with the decomposition
\begin{eqnarray}
  \chi_{\di_\ell}^{so(2,d)}(q^2, \bar x^2) & = & \epsilon_r
  \sum_{t=-\ell+1}^{\ell-1} \chi^{so(2,d)}_{[d-2t-1;0]}(q,\bar x) +
  \epsilon_{r-1} \sum_{t=1,3,\dots}^{2\ell-3} \Big[
    \chi^{so(2,d)}_{[d+t-1;0]}(q, \bar x) -
    \chi^{so(2,d)}_{[d-t-1;0]}(q, \bar x) \Big] \nonumber \\ && \qquad
  \quad -\sum_{m=0}^{r-1} \epsilon_m \sum_{t=1,3,\dots}^{2\ell-1}
  \sum_{s=1}^\infty (-1)^s \chi^{so(2,d)}_{[s+d-t-1;s,1^{r-1-m}]}(q,
  \bar x) \nonumber \\ && \qquad \qquad - \sum_{m=0}^{r-1} \epsilon_m
  \sum_{t=1,3,\dots}^{2\ell-3} \sum_{s=1}^\infty (-1)^s
  \chi^{so(2,d)}_{[s+d-t-1;s,1^{r-1-m}]}(q, \bar x)\,.
\end{eqnarray}
Due to the presence of the alternating signs $\epsilon_m$ in the above
expression, the spectrum of the minimanl type-B$_\ell$ model depends
on the parity of the integer part of the rank $r$. Introducing
\begin{equation}
  \bigoplus_{t\,\,{\rm odd}} := \bigoplus_{t=1,3,\dots}^{2\ell-1}
  \oplus \bigoplus_{t=1,3,\dots}^{2\ell-3}\,,
\end{equation}
the four possible cases read as follow:
\begin{itemize}
\item \underline{Even rank $r=2n$ with $n=2p$:}
  \begin{eqnarray}
    \di_\ell^{\wedge 2} & \cong & \bigoplus_{t=1,3,\dots}^{2\ell-3}
    \mathcal{D}\big(d-t-1;0\big) \,\,\oplus\,\, \bigoplus_{t=2}^{2\ell-2}\,\,
    \bigoplus_{m=0}^{r-1}\,\, \bigoplus_{s=1}^\infty \mathcal{D}\big(
    s+d-t-1;s,1^m\big)
     \nn && \qquad \oplus \bigoplus_{t\,\, {\rm odd}}\,\,
    \bigoplus_{m=0,3\,\,{\rm mod} 4}\,\,
    \bigoplus_{s=2,4,\dots}^\infty \mathcal{D}\big( s+d-t-1;s,1^m\big)
     \\ && \qquad \oplus \bigoplus_{t\,\,
      {\rm odd}}\,\, \bigoplus_{m=1,2\,\,{\rm mod} 4}\,\,
    \bigoplus_{s=1,3,\dots}^\infty \mathcal{D}\big( s+d-t-1;s,1^m\big)\,; \nonumber
  \end{eqnarray}
\item \underline{Even rank $r=2n$ with $n=2p+1$:}
  \begin{eqnarray}
    \di_\ell^{\wedge 2} & \cong & \bigoplus_{t=-\ell+1}^{\ell-1}
    \mathcal{D}\big(d-2t-1;0\big) \oplus
    \bigoplus_{t=1,3,\dots}^{2\ell-3} \mathcal{D}\big(d+t-1;0\big)
     \nn && \qquad \oplus \bigoplus_{t\,\, {\rm odd}}\,\,
    \bigoplus_{m=1,2\,\,{\rm mod}\, 4}\,\,
    \bigoplus_{s=2,4,\dots}^\infty \mathcal{D}\big( s+d-t-1;s,1^m\big)
     \nn && \qquad \quad \oplus \bigoplus_{t\,\, {\rm odd}}\,\,
    \bigoplus_{m=0,3\,\,{\rm mod}\, 4}\,\,
    \bigoplus_{s=1,3,\dots}^\infty \mathcal{D}\big( s+d-t-1;s,1^m\big)
     \nn && \qquad \qquad \quad \oplus \bigoplus_{t=2}^{2\ell-2}\,\,
    \bigoplus_{m=0}^{r-1}\,\, \bigoplus_{s=1}^\infty \mathcal{D}\big(
    s+d-t-1;s,1^m\big)\,;
  \end{eqnarray}
\item \underline{Even rank $r=2n+1$ with $n=2p$:}
  \begin{eqnarray}
    \di_\ell^{\wedge 2} & \cong & \bigoplus_{t=-\ell+1}^{\ell-1}
    \mathcal{D}\big(d-2t-1;0\big) \oplus
    \bigoplus_{t=1,3,\dots}^{2\ell-3} \mathcal{D}\big(d-t-1;0\big)
     \nn && \qquad \oplus \bigoplus_{t\,\, {\rm odd}}\,\,
    \bigoplus_{m=0,1\,\,{\rm mod}\, 4}\,\,
    \bigoplus_{s=2,4,\dots}^\infty \mathcal{D}\big( s+d-t-1;s,1^m\big)
     \nn && \qquad \quad \oplus \bigoplus_{t\,\, {\rm odd}}\,\,
    \bigoplus_{m=2,3\,\,{\rm mod}\, 4}\,\,
    \bigoplus_{s=1,3,\dots}^\infty \mathcal{D}\big( s+d-t-1;s,1^m\big)
     \nn && \qquad \qquad \quad \oplus \bigoplus_{t=2}^{2\ell-2}\,\,
    \bigoplus_{m=0}^{r-1}\,\, \bigoplus_{s=1}^\infty \mathcal{D}\big(
    s+d-t-1;s,1^m\big)\,;
  \end{eqnarray}
\item \underline{Even rank $r=2n+1$ with $n=2p+1$:}
  \begin{eqnarray}
    \di_\ell^{\wedge 2} & \cong & \bigoplus_{t=1,3,\dots}^{2\ell-3}
    \mathcal{D}\big(d+t-1;0\big) \oplus \bigoplus_{t=2}^{2\ell-2}\,\,
    \bigoplus_{m=0}^{r-1}\,\, \bigoplus_{s=1}^\infty \mathcal{D}\big(
    s+d-t-1;s,1^m\big) \nn && \qquad \oplus \bigoplus_{t\,\,
      {\rm odd}}\,\, \bigoplus_{m=0,1\,\,{\rm mod} 4}\,\,
    \bigoplus_{s=1,3,\dots}^\infty \mathcal{D}\big( s+d-t-1;s,1^m\big)
     \\ && \qquad \oplus \bigoplus_{t\,\, {\rm odd}}\,\,
    \bigoplus_{m=2,3\,\,{\rm mod} 4}\,\,
    \bigoplus_{s=2,4,\dots}^\infty \mathcal{D}\big( s+d-t-1;s,1^m\big)\,.\nonumber
  \end{eqnarray}
\end{itemize}

\subsection{Even $d$}
\label{app:Bleven}
As recalled previously, in odd-dimensional AdS space (i.e. when
$d=2r$) one can consider a chiral $\di_\ell$ singleton, that is, a Weyl
spinor subject to a higher-order Dirac equation on the
$d$-dimensional conformal boundary. As a consequence, a first
truncation, before the minimal model, of the type-B$_\ell$ theory is 
the chiral type-B$_{\ell,\pm}$ whose spectrum is given by
the tensor product of two $\di_\ell$ singleton of same chirality. This
decomposition is presented hereafter.

\subsubsection{Chiral Flato-Fronsdal}
\label{app:chiralFF}
The main ingredient in obtaining the decomposition displayed below is
the $so(2) \oplus so(d)$ decomposition of a chiral singleton module,
namely
\begin{equation}
  \chi^{so(d)}_{\di_{\ell,\pm}}(q, \bar x) = \sum_{s=0}^\infty
  q^{\frac{d+1-2\ell}2+s}\, \Big( \sum_{k=0}^{\ell-1} q^{2k}\,
  \chi_{(s+\frac12,\frac12_\pm)}^{so(d)}(\bar x)\, +
  \sum_{k=0}^{\ell-2} q^{2k+1}\,
  \chi^{so(d)}_{(s+\frac12,\frac12_\mp)}(\bar x) \Big)\, ,
\end{equation}
where the second term appears only for $\ell>1$. With this identity at
hand, one can show that
\begin{itemize}
\item \underline{Rank $r=2n$:} The tensor product of two $\di_\ell$
  singleton of the same chirality decomposes as follows:
  \begin{eqnarray}
    \di_{\ell,\pm}^{\otimes 2} & = &
    \bigoplus_{t=-2\ell+2,-2\ell+4,\dots}^{2\ell-2}
    \mathcal{D}\big(d-t-1;0\big) \oplus \bigoplus_{m=1}^{n}
    \bigoplus_{t=1,3,\dots}^{2\ell-1} \bigoplus_{s=1}^{\infty}
    \mathcal{D}\big(s+d-t-1;s,1_\pm^{2m-1}\big)  \nn && \qquad
    \oplus \bigoplus_{m=1}^{n} \bigoplus_{t=1,3,\dots}^{2\ell-3}
    \bigoplus_{s=1}^{\infty}
    \mathcal{D}\big(s+d-t-1;s,1_\mp^{2m-1}\big)  \nn && \qquad \qquad
    \qquad \oplus 2\,\bigoplus_{m=0}^{n-1}
    \bigoplus_{t=2,4,\dots}^{2\ell-2} \bigoplus_{s=1}^{\infty}
    \mathcal{D}\big(s+d-t-1;s,1^{2m}\big)\,.  
    \label{chiralFFBleven}
  \end{eqnarray}
  Notice in particular that there is no graviton in this spectrum,
  but the massive scalars are present. The tensor product of two
  opposite chiralities $\di_\ell$ singleton reads
  \begin{eqnarray}
    \di_{\ell,+} \otimes \di_{\ell,-} & = &
    \bigoplus_{t=-2\ell+3,-2\ell+5,\dots}^{2\ell-3}
    \mathcal{D}\big(d-t-1;0\big) \oplus \bigoplus_{m=1}^{n}
    \bigoplus_{s=1}^{\infty} \bigoplus_{t=2,4,\dots}^{2\ell-2}
    \mathcal{D}\big(s+d-t-1;s,1_\pm^{2m-1}\big) \nn && \qquad
    \oplus \bigoplus_{m=1}^{n} \bigoplus_{s=1}^{\infty}
    \bigoplus_{t=2,4,\dots}^{2\ell-2}
    \mathcal{D}\big(s+d-t-1;s,1_\mp^{2m-1}\big) \nn&& \qquad \qquad
    \oplus \bigoplus_{m=0}^{n-1} \bigoplus_{s=1}^\infty
    \Big(\bigoplus_{t=1,3,\dots}^{2\ell-1} \oplus
    \bigoplus_{t=1,3,\dots}^{2\ell-3} \Big)
    \mathcal{D}\big(s+d-t-1;s,1^{2m}\big)\,. 
   \end{eqnarray}
\item \underline{Rank $r=2n+1$:} Contrary to the previous case, the
  tensor product of two $\di_\ell$ singletons of the same chirality which
  reads
  \begin{eqnarray}
    \di_{\ell,\pm}^{\otimes 2} & = &
    \bigoplus_{t=-2\ell+3,-2\ell+5,\dots}^{2\ell-3}
    \mathcal{D}\big(d-t-1;0\big) \oplus \bigoplus_{m=0}^{n}
    \bigoplus_{t=1,3,\dots}^{2\ell-1} \bigoplus_{s=1}^{\infty}
    \mathcal{D}\big(s+d-t-1;s,1^{2m}_\pm\big) \nn && \qquad
    \quad \oplus \bigoplus_{m=0}^n \bigoplus_{t=1,3,\dots}^{2\ell-3}
    \mathcal{D}\big(s+d-t-1;s,1^{2m}_\mp\big) \label{chiralFFBlodd}
    \nn && \qquad \qquad \qquad \oplus 2\bigoplus_{m=1}^{n}
    \bigoplus_{s=1}^{\infty} \bigoplus_{t=2,4,\dots}^{2\ell-2}
    \mathcal{D}\big(s+d-t-1;s,1^{2m-1}\big)\,.
  \end{eqnarray}
  The above does contain a graviton but no longer include the massive
  scalars. The tensor product of two $\di_\ell$ singletons of opposite
  chiralities displays the complement of the previous content with
  respect to the spectrum of the (full) type-B$_\ell$ theory:
  \begin{eqnarray}
    \di_{\ell,+} \otimes \di_{\ell,-} & = & \bigoplus_{t=-2\ell+2,
      -2\ell+4,\dots}^{2\ell-2} \mathcal{D}\big(d-t-1;0\big) \oplus
    \bigoplus_{m=0}^{n} \bigoplus_{s=1}^{\infty}
    \bigoplus_{t=2,4,\dots}^{2\ell-2}
    \mathcal{D}\big(s+d-t-1;s,1^{2m}_\pm\big) \nonumber \\ && \qquad
    \oplus \bigoplus_{m=0}^{n} \bigoplus_{s=1}^{\infty}
    \bigoplus_{t=2,4,\dots}^{2\ell-2}
    \mathcal{D}\big(s+d-t-1;s,1^{2m}_\mp\big) \nn && \qquad \qquad
    \oplus \bigoplus_{m=1}^{n} \bigoplus_{s=1}^{\infty} \Big(
    \bigoplus_{t=1,3,\dots}^{2\ell-1} \oplus
    \bigoplus_{t=1,3,\dots}^{2\ell-3} \Big) \mathcal{D}\big(
    s+d-t-1;s,1^{2m-1}\big)\,.
  \end{eqnarray}
\end{itemize}

\subsubsection{Minimal model}
\label{app:minBleven}
In order to decompose $\chi^{so(2,d)}_{\di_{\ell,\pm}}(q^2, \bar x^2)$
into a sum of characters appearing in \eqref{chiralFFBleven} and
\eqref{chiralFFBlodd}, we need the identity
\begin{equation}
  \chi^{so(d)}_{\frac12_\pm}(\bar x^2) = \sum_{m=0}^{[r/2]} (-1)^m\,
  \chi^{so(d)}_{(1^{r-2m}_\pm)}(\bar x)\, .
\end{equation}
Using the above property, one can show that for $r=2n$,
\begin{eqnarray}
  \chi^{so(2,d)}_{\di_{\ell,\pm}}(q^2, \bar x^2) & = & (-1)^n
  \sum_{t=-2\ell+2,-2\ell+4,\dots}^{2\ell-2}
  \chi^{so(2,d)}_{[d-t-1;0]}(q, \bar x)  \nn && \qquad + \sum_{m=1}^n
  (-1)^{n+m+1} \sum_{s=1}^\infty (-1)^s \Big[
    \sum_{t=1,3,\dots}^{2\ell-1}
    \chi^{so(2,d)}_{[s+d-t-1;s,1^{2m-1}_\pm]}(q, \bar x)  \nn && \qquad
    \qquad \quad + \sum_{t=1,3,\dots}^{2\ell-3}
    \chi^{so(2,d)}_{[s+d-t-1;s,1^{2m-1}_\mp]}(q, \bar x) \Big]\,,
\end{eqnarray}
whereas for $r=2n+1$,
\begin{eqnarray}
  \chi^{so(2,d)}_{\di_{\ell,\pm}}(q^2, \bar x^2) & = & (-1)^n
  \sum_{t=1,3,\dots}^{2\ell-3} \Big[ \chi^{so(2,d)}_{[d+t-1;0]}(q,
    \bar x) - \chi^{so(2,d)}_{[d-t-1;0]}(q, \bar x) \Big]  \nn && \qquad
  +\sum_{m=0}^n (-1)^{n+m+1} \sum_{s=1}^\infty (-1)^s \Big[
    \sum_{t=1,3,\dots}^{2\ell-1}
    \chi^{so(2,d)}_{[s+d-t-1;s,1^{2m}_\pm]}(q, \bar x)  \nn && \qquad
    \qquad \quad + \sum_{t=1,3,\dots}^{2\ell-3}
    \chi^{so(2,d)}_{[s+d-t-1;s,1^{2m}_\mp]}(q, \bar x) \Big]\,,
\end{eqnarray}
and hence
\begin{itemize}
\item \underline{Even rank $r=2n$ with $n=2p$:}
  \begin{eqnarray} 
    \di_{\ell,\pm}^{\wedge 2} & \cong & \bigoplus_{m=3\,\,{\rm
        mod}\,\,4}\,\, \bigoplus_{s=2,4,\dots}^{\infty}
    \bigoplus_{t=1,3,\dots}^{2\ell-1}
    \mathcal{D}\big(s+d-t-1;s,1_\pm^{m}\big) \\ && \qquad \oplus
    \bigoplus_{m=3\,\,{\rm mod}\,\,4}\,\,
    \bigoplus_{s=2,4,\dots}^{\infty} \bigoplus_{t=1,3,\dots}^{2\ell-3}
    \mathcal{D}\big(s+d-t-1;s,1_\mp^{m}\big) \nonumber \\ && \qquad
    \qquad \oplus \bigoplus_{m=1\,\,{\rm mod}\,\,4}\,\,
    \bigoplus_{t=1,3,\dots}^{2\ell-1} \bigoplus_{s=1,3,\dots}^{\infty}
    \mathcal{D}\big(s+d-t-1;s,1_\pm^{m}\big) \nonumber \\ && \qquad
    \qquad \qquad \oplus \bigoplus_{m=1\,\,{\rm mod}\,\,4}\,\,
    \bigoplus_{s=1,3,\dots}^{\infty} \bigoplus_{t=1,3,\dots}^{2\ell-3}
    \mathcal{D}\big(s+d-t-1;s,1_\mp^{m}\big) \nonumber \\ && \qquad
    \qquad \qquad \qquad \oplus \bigoplus_{m=0}^{n-1}\,\,
    \bigoplus_{s=1}^{\infty} \bigoplus_{t=2,4,\dots}^{2\ell-2}
    \mathcal{D}\big(s+d-t-1;s,1^{2m}\big)\,; \nonumber
  \end{eqnarray}
\item \underline{Even rank $r=2n$ with $n=2p+1$:}
  \begin{eqnarray}
    \di_{\ell,\pm}^{\wedge 2} & \cong &
    \bigoplus_{t=-2\ell+2,-2\ell+4,\dots}^{2\ell-2}
    \mathcal{D}\big(d-t-1;0\big) \\ && \qquad \oplus
    \bigoplus_{m=1\,\,{\rm mod}\,\,4}\,\,
    \bigoplus_{s=2,4,\dots}^{\infty} \bigoplus_{t=1,3,\dots}^{2\ell-1}
    \mathcal{D}\big(s+d-t-1;s,1_\pm^{m}\big) \nonumber \\ && \qquad
    \qquad \oplus \bigoplus_{m=1\,\,{\rm mod}\,\,4}\,\,
    \bigoplus_{s=2,4,\dots}^{\infty} \bigoplus_{t=1,3,\dots}^{2\ell-3}
    \mathcal{D}\big(s+d-t-1;s,1_\mp^{m}\big) \nonumber \\ && \qquad
    \qquad \qquad \oplus \bigoplus_{m=3\,\,{\rm mod}\,\,4}\,\,
    \bigoplus_{s=1,3,\dots}^{\infty} \bigoplus_{t=1,3,\dots}^{2\ell-1}
    \mathcal{D}\big(s+d-t-1;s,1_\pm^{m}\big) \nonumber \\ && \qquad
    \qquad \qquad \qquad \oplus \bigoplus_{m=3\,\,{\rm mod}\,\,4}\,\,
    \bigoplus_{s=1,3,\dots}^{\infty} \bigoplus_{t=1,3,\dots}^{2\ell-3}
    \mathcal{D}\big(s+d-t-1;s,1_\mp^{m}\big) \nonumber \\ && \qquad
    \qquad \qquad \qquad \qquad \oplus \bigoplus_{m=0}^{n-1}\,\,
    \bigoplus_{t=2,4,\dots}^{2\ell-2} \bigoplus_{s=1}^{\infty}
    \mathcal{D}\big(s+d-t-1;s,1^{2m}\big)\,;\nonumber
  \end{eqnarray}
\item \underline{Even rank $r=2n+1$ with $n=2p$:}
  \begin{eqnarray}
    \di_{\ell,\pm}^{\wedge 2} & \cong & \bigoplus_{t=1,3,\dots}^{2\ell-3}
    \mathcal{D}\big(d-t-1;0\big) \\ && \qquad \oplus
    \bigoplus_{m=0\,\,{\rm mod}\,\,4} \bigoplus_{s=2,4,\dots}^{\infty}
    \bigoplus_{t=1,3,\dots}^{2\ell-1}
    \mathcal{D}\big(s+d-t-1;s,1^{m}_\pm\big) \nonumber \\ && \qquad
    \qquad \oplus \bigoplus_{m=0\,\,{\rm mod}\,\,4}
    \bigoplus_{s=2,4,\dots}^{\infty} \bigoplus_{t=1,3,\dots}^{2\ell-3}
    \mathcal{D}\big(s+d-t-1;s,1^{m}_\mp\big) \nonumber \\ && \qquad
    \qquad \qquad \oplus \bigoplus_{m=2\,\,{\rm mod}\,\,4}
    \bigoplus_{s=1,3,\dots}^{\infty} \bigoplus_{t=1,3,\dots}^{2\ell-1}
    \mathcal{D}\big(s+d-t-1;s,1^{m}_\pm\big) \nonumber \\ && \qquad
    \qquad \qquad \qquad \oplus \bigoplus_{m=0\,\,{\rm mod}\,\,4}
    \bigoplus_{s=1,3,\dots}^{\infty} \bigoplus_{t=1,3,\dots}^{2\ell-3}
    \mathcal{D}\big(s+d-t-1;s,1^{m}_\mp\big) \nonumber \\ && \qquad
    \qquad \qquad \qquad \quad \oplus \bigoplus_{m=1}^{n}
    \bigoplus_{s=1}^{\infty} \bigoplus_{t=2,4,\dots}^{2\ell-2}
    \mathcal{D}\big(s+d-t-1;s,1^{2m-1}\big)\,; \nonumber
  \end{eqnarray}
\item \underline{Even rank $r=2n+1$ with $n=2p+1$:}
  \begin{eqnarray}
    \di_{\ell,\pm}^{\wedge 2} & \cong & \bigoplus_{t=1,3,\dots}^{2\ell-3}
    \mathcal{D}\big(d+t-1;0\big) \\ && \qquad \oplus
    \bigoplus_{m=0\,\,{\rm mod}\,\,4}\,\,
    \bigoplus_{s=1,3,\dots}^{\infty} \bigoplus_{t=1,3,\dots}^{2\ell-1}
    \mathcal{D}\big(s+d-t-1;s,1^{2m}_\pm\big) \nonumber \\ && \qquad
    \qquad \oplus \bigoplus_{m=0\,\,{\rm mod}\,\,4}\,\,
    \bigoplus_{s=1,3,\dots}^{\infty} \bigoplus_{t=1,3,\dots}^{2\ell-3}
    \mathcal{D}\big(s+d-t-1;s,1^{2m}_\mp\big) \nonumber \\ && \qquad
    \qquad \qquad \oplus \bigoplus_{m=2\,\,{\rm mod}\,\,4}\,\,
    \bigoplus_{s=2,4,\dots}^{\infty} \bigoplus_{t=1,3,\dots}^{2\ell-1}
    \mathcal{D}\big(s+d-t-1;s,1^{2m}_\pm\big) \nonumber \\ && \qquad
    \qquad \qquad \qquad \oplus \bigoplus_{m=2\,\,{\rm mod}\,\,4}\,\,
    \bigoplus_{s=2,4,\dots}^{\infty} \bigoplus_{t=1,3,\dots}^{2\ell-3}
    \mathcal{D}\big(s+d-t-1;s,1^{2m}_\mp\big) \nonumber \\ && \qquad
    \qquad \qquad \qquad \qquad \oplus \bigoplus_{m=1}^{n}
    \bigoplus_{s=1}^{\infty} \bigoplus_{t=2,4,\dots}^{2\ell-2}
    \mathcal{D}\big(s+d-t-1;s,1^{2m-1}\big)\,. \nonumber
  \end{eqnarray}
\end{itemize}

\newpage
\bibliographystyle{JHEP}
\bibliography{biblio_ii}

\end{document}